\RequirePackage{fix-cm}
\documentclass[twocolumn, epjc3, comma, sort&compress, natbib]{svjour3}
\bibpunct{[}{]}{,}{n}{}{,} 

\journalname{Eur. Phys. J.}

\usepackage{latexsym}
\usepackage{amsmath}
\usepackage{amssymb}
\usepackage{amsfonts}
\usepackage{CJKutf8}
\usepackage{slashed}

\usepackage[mathscr,scaled=1.15]{urwchancal}
\DeclareFontFamily{OT1}{pzc}{}
\DeclareFontShape{OT1}{pzc}{m}{it}%
{<-> s * [1.15] pzcmi7t}{}
\DeclareMathAlphabet{\mathpzc}{OT1}{pzc}{m}{it}

\usepackage{color}

\usepackage{supertabular}
\usepackage{placeins}
\usepackage{epsfig}
\usepackage{graphicx}

\definecolor{purple}{rgb}{0.5,0,0.5}
\definecolor{blue}{rgb}{0.0,0,0.9}
\definecolor{prdblue}{rgb}{0.133,0.118,0.498}
\usepackage[colorlinks=true, pdfstartview=FitV, linkcolor=prdblue, citecolor= prdblue, urlcolor=prdblue]{hyperref}

\begin{document}

\begin{CJK}{UTF8}{song}

\title{$\,$\\[-6ex]\hspace*{\fill}{\normalsize{\sf\emph{Preprint nos}.\
USTC-ICTS/PCFT-23-14,
NJU-INP 074/23
}
}\\[1ex]
$\Delta$-baryon axialvector and pseudoscalar form factors, \\ and associated PCAC relations}

\author{
Pei-Lin Yin\thanksref{NJPT}%
        $\,^{\href{https://orcid.org/0000-0001-7198-8157}{\textcolor[rgb]{0.00,1.00,0.00}{\sf ID}}}$
        \and
%
    Chen Chen\thanksref{USTC1,USTC2}%
        $\,^{\href{https://orcid.org/0000-0003-3619-0670}{\textcolor[rgb]{0.00,1.00,0.00}{\sf ID}}}$
    \and \\
    Christian S. Fischer\thanksref{UG1,UG2}%
    $\,^{\href{https://orcid.org/0000-0001-8780-7031}{\textcolor[rgb]{0.00,1.00,0.00}{\sf ID}}}$
    \and
       Craig D.~Roberts\thanksref{NJU,INP}%
       $\,^{\href{https://orcid.org/0000-0002-2937-1361}{\textcolor[rgb]{0.00,1.00,0.00}{\sf ID}}}$
}

\authorrunning{P.-L. Yin, C.~Chen, C.\,S.~Fischer and C.\,D.~Roberts} 

\institute{College of Science, Nanjing University of Posts and Telecommunications, Nanjing 210023, China \label{NJPT}
\and
Interdisciplinary Center for Theoretical Study, University of Science and Technology of China, Hefei, Anhui 230026, China \label{USTC1}
            \and
            Peng Huanwu Center for Fundamental Theory, Hefei, Anhui 230026, China \label{USTC2}
\and
Institut f\"ur Theoretische Physik, Justus-Liebig-Universit\"at Gie{\ss}en, D-35392 Gie{\ss}en, Germany \label{UG1}
\and
Helmholtz Forschungsakademie Hessen f\"ur FAIR (HFHF),
GSI Helmholtzzentrum f\"ur Schwerionenforschung, \\ \hspace*{1em}Campus Gie{\ss}en, 35392 Gie{\ss}en, Germany \label{UG2}
            \and
            School of Physics, Nanjing University, Nanjing, Jiangsu 210093, China \label{NJU}
           \and
           Institute for Nonperturbative Physics, Nanjing University, Nanjing, Jiangsu 210093, China \label{INP}
\\[1ex]
Email:
\href{mailto:chenchen1031@ustc.edu.cn}{chenchen1031@ustc.edu.cn} (C.~Chen);
\href{mailto:cdroberts@nju.edu.cn}{cdroberts@nju.edu.cn} (C.\,D.~Roberts).
            }


\date{2023 May 16}

\maketitle

\end{CJK}

\begin{small}
\begin{abstract}
A quark+diquark Faddeev equation treatment of the baryon bound state problem in Poincar\'e-invariant quantum field theory is used to deliver para\-meter-free predictions for all six $\Delta$-baryon elastic weak form factors.
Amongst the results, it is worth highlighting that there are two distinct classes of such $\Delta$-baryon form factors, $(g_1, g_3, G_{\pi\Delta\Delta})$,  $(h_1, h_3, H_{\pi\Delta\Delta})$, the functions within each of which are separately linked via partial conservation of axial current (PCAC) and Goldber\-ger-Treiman (GT) relations.
Respectively with\-in each class, the listed form factors possess qualitatively the same structural features as the nucleon axial, induced pseudoscalar, and pion-nucleon coupling form factors.
For instance, the $\Delta$-baryon $g_1$ axial form factor can reliably be approximated by a dipole function, characterised by an axial charge $g_A^{\Delta^+}=0.71(9)$ and mass-scale $m_A^\Delta=0.95(2)m_\Delta$.
Moreover, the two distinct $\Delta$-baryon PCAC form factor relations are satisfied to a high degree of accuracy on a large range of $Q^2$; the associated GT relations present good approximations only on $Q^2/m_\Delta^2 \simeq 0$;
and pion pole dominance approximations are reliable within both classes.
There are two $\pi\Delta\Delta$ couplings: $g_{\pi\Delta\Delta} = 10.46(1.88)$; $h_{\pi\Delta\Delta}= 35.73(3.75)$; and the associated form factors are soft.   Such couplings commonly arise in phenomenology, which may therefore benefit from our analyses.
A flavour decomposition of the axial charges reveals that quarks carry $71$\% of the $\Delta$-baryon spin.  The analogous result for the proton is $\approx 65$\%.
\end{abstract}
\end{small}


\section{Introduction}
\label{secintro}
The response of baryons to electromagnetic probes is much studied, both experimentally \cite{Holt:2010vj, Holt:2012gg, Punjabi:2015bba, Brodsky:2015aia, Carman:2020qmb, Anderle:2021wcy, AbdulKhalek:2021gbh, Quintans:2022utc, Carman:2023zke} and theoretically \cite{Eichmann:2016yit, Brodsky:2020vco, Ding:2022ows}.  An entirely new perspective on baryon structure is provided by weak-interaction probes, with form factors that can be measured in, \emph{e.g}., neutrino-nucleus scattering.  Here, nucleon axialvector and pseudoscalar form factors are the archetypes, being crucial inputs for Standard Model tests via weak interactions, neutrino-nucleus scattering and parity violation experiments.  Consequently, a diverse array of theory tools -- using both continuum \cite{Anikin:2016teg, Chen:2020wuq, Chen:2021guo, ChenChen:2022qpy} and lattice \cite{Alexandrou:2017hac, Jang:2019vkm, Bali:2019yiy} formulations of hadron bound state problems -- has recently been employed to deliver a better understanding of their behaviour.

The lightest excitations of the nucleon are the \linebreak $\Delta(1232)$-baryons.  Theoretically, as $(I,J^P)=(\tfrac{3}{2},\tfrac{3}{2}^+)$ systems, $\Delta$-baryons are less complex than $(\tfrac{1}{2},\tfrac{1}{2}^+)$ nucleons because their Poincar\'e-covariant wave functions are simpler.  For instance, viewed from a modern quark + diquark perspective \cite{Barabanov:2020jvn}, $\Delta$-baryons only contain iso\-vector-axialvector diquark correlations \cite{Liu:2022ndb}, whereas iso\-scalar-scalar diquarks also play a large role in nucleons.  Such structural distinctions make comparisons between predictions for nucleon and $\Delta$-baryon properties useful in developing an understanding of how quantum chromodynamics (QCD) produces systems constituted from three valence quarks.  These features explain why much theoretical attention has been devoted to the calculation of $\Delta$-baryon elastic electromagnetic form factors \cite{Alexandrou:2007we, Alexandrou:2009hs, Nicmorus:2010sd, Ledwig:2011cx, Alexandrou:2012da, Segovia:2013uga, Segovia:2014aza, Kim:2019gka}, even though measurement of such form factors is impossible because of the very short lifetime of these resonances: $\tau_{\Delta} \approx 10^{-26} \tau_{n}$, where $\tau_{n}$ is the lifetime of a free neutron.  (Estimates of $\Delta$-baryon magnetic moments have been produced through analyses of $\pi^+ p \to \pi^+ p \gamma$ reactions \cite[RPP]{Workman:2022ynf}.)

Against this backdrop, it is natural to develop comparative studies of the weak-interaction structure of the nucleon and $\Delta$-baryon.  As well as being interesting in their own right, predictions for such quantities as the $\Delta$-baryon axial charge, $g_A^\Delta$, and $\pi\Delta\Delta$ coupling, unified with analogous nucleon properties, can provide valuable inputs (constraints) for effective field theories employed in low-energy hadron physics \cite{Jiang:2018mzd, Holmberg:2018dtv}.  The calculation of $\Delta$-baryon elastic weak-interaction form factors is also a useful preliminary to delivering reliable predictions for the weak-probe induced $N\to \Delta$ transition, whose form factors are experimentally accessible and which may play an important role in understanding long-baseline and atmospheric neutrino-nucleus scattering experiments \cite{Mosel:2016cwa, Alvarez-Ruso:2017oui, Lovato:2020kba, Simons:2022ltq}.

It is thus unsurprising that numerous analyses have computed values for $g_A^\Delta$ and $g_{\pi\Delta\Delta}$ -- see, \emph{e.g}., Refs.\,\cite{Brown:1975di, Dashen:1993jt, Zhu:2000zd, Alexandrou:2011py, Alexandrou:2013opa, Alexandrou:2016xok, Jiang:2008we, Choi:2010ty, Buchmann:2013fxa, Kucukarslan:2014bla, Wang:2015osq, Yao:2016vbz, Liu:2018jiu, Jun:2020lfx}.   Amongst them, however, only one pair of studies \cite{Alexandrou:2011py, Alexandrou:2013opa}, working with lattice-regularised QCD (lQCD), has provided results for the entire set of four-plus-two form factors required to completely describe $\Delta$-baryon axial and pseudoscalar currents.  Unfortunately, those calculations were performed with unphysically large pion masses and the results possess significant uncertainties.  A chiral quark-soliton model ($\chi$QSM) has been used to compute the four axial form factors \cite{Jun:2020lfx}.

Continuum Schwinger function methods (CSMs) provide an alternative to models and lQCD computations in hadron physics.  In such applications, contemporary progress and challenges are canvassed elsewhere \cite{Eichmann:2016yit, Fischer:2018sdj, Qin:2020rad, Brodsky:2020vco, Binosi:2022djx, Papavassiliou:2022wrb, Ding:2022ows, Mezrag:2023nkp, Ferreira:2023fva}.  Of particular relevance herein is the recent construction and use \cite{Chen:2020wuq, Chen:2021guo, ChenChen:2022qpy} of symmetry-preserving axial and pseudoscalar currents appropriate for baryons described by the fully-interac\-ting quark+nonpointlike-diquark Faddeev equation introduced in Refs.\,\cite{Cahill:1988dx, Reinhardt:1989rw, Efimov:1990uz}.  This has enabled the use of CSMs to complete a parame\-ter-free comparative study and unification of the weak-interaction structure of the nucleon and $\Delta$-baryon.  At the simplest level, the outcomes can be used to test the current construction via comparisons with results from models and lQCD.  Passing such tests, sound predictions for weak $N\to \Delta$ transitions can follow.
Such predictions can be tested because, \emph{e.g}., data exist \cite{CLAS:2017fja, CLAS:2018fon} from which the axial $\pi N \to \Delta$ transition form factor may be extracted after extending existing reaction models \cite{Mokeev:2008iw}.

Our discussion is organised as follows.  Section~\ref{secdef} explains the structure of the $\Delta$-baryon elastic matrix elements of the axialvector and pseudoscalar currents and introduces the full array of associated form factors.  The quark+diquark Faddeev equation used to describe the $\Delta$-baryon is sketched in Sec.\,\ref{sectheo} along with the related symmetry preserving current.  Our results are presented in Sec.\,\ref{secnum}, wherein they are also compared with those from other studies, and followed in Sec.\,\ref{SecFlavourSep} with a brief discussion of the flavour-separated $\Delta$-baryon axial charges and the quark contribution to their total spin.
Section~\ref{secsum} contains a summary and perspective.  Numerous technical details are collected in appendices.


\section{$\Delta(1232)$ axial and pseudoscalar currents}
\subsection{General structure}
\label{secdef}
Introducing the column vector $\psi=(u,d)^{\rm T}$, where $u$, $d$ are quark fields, the axialvector current operator can be written ${\mathcal A}^{j}_{5\mu}(x) = \bar{\psi}(x)\frac{\tau^j}{2}\gamma_5\gamma_\mu \psi(x)$, where the isospin (flavour) structure is given by the Pauli matrices, $\{\tau^i|i=1,2,3\}$, with $\tau^3$ representing the neutral current and $\tau^{1\pm i2}:=(\tau^1\pm i\tau^2)/2$ expressing the charged currents.  The in-$\Delta$ expectation value of this operator is \cite{Alexandrou:2011py, Alexandrou:2013opa}:
\begin{subequations}
\label{axdef}
\begin{align}
{\cal J}_{5\mu}(K,Q):=
\langle\Delta(P_f;s_f)|{\mathcal A}^{j}_{5\mu}(0)|\Delta(P_i;s_i)\rangle\\
=\bar{u}_\alpha(P_f;s_f)	\Gamma_{5\mu,\alpha\beta}(Q)u_\beta(P_i;s_i)\,, \label{RSspinor}
\end{align}
\end{subequations}
where $P_i(s_i)$ and $P_f(s_f)$ are $\Delta$-baryon incoming and outgoing momenta (spins), with $P_f^2 = -m_\Delta^2 = P_i^2 $;
and
$K=(P_f+P_i)/2$ is the average momentum and $Q=P_f-P_i$ is the momentum of the weak probe.
%
In writing Eq.\,\eqref{RSspinor}, we have used a Euclidean space Rarita-Schwinger spinor, which is the same for all $\Delta$-baryons and whose properties are explained elsewhere \cite[Appendix\,B]{Segovia:2014aza}.
Notably, the choice of $j$ constrains the initial and final charge-states of the $\Delta$-baryon, \emph{e.g}.,  $j=3$ entails that they are both the same.

The $\Delta$-baryon axial current vertex in Eq.\,\eqref{RSspinor} has the general form:
\begin{align}
\label{axffsdef}
\nonumber
\Gamma_{5\mu,\alpha\beta}& (Q) = -\frac{1}{2}\gamma_5\bigg[
\delta_{\alpha\beta}\bigg(g_1(Q^2)\gamma_\mu + ig_3(Q^2)\frac{Q_\mu}{2m_\Delta}\bigg)\\
& -\frac{Q_\alpha Q_\beta}{4m_\Delta^2}\bigg(h_1(Q^2)\gamma_\mu + ih_3(Q^2)\frac{Q_\mu}{2m_\Delta}\bigg)\bigg]\,,
\end{align}
where $g_1$, $g_3$, $h_1$, $h_3$ are four Poincar\'e invariant form factors.

The matrix element of the analogous pseudoscalar operator, ${\mathpzc P}^j_{5}(x)$, is
\begin{align}
{\cal J}_{5}(K,Q):=
\bar{u}_\alpha(P_f;s_f)	\Gamma_{5,\alpha\beta}(Q)u_\beta(P_i;s_i)\,,
\label{psdef}
\end{align}
with
\begin{align}
\label{psffsdef}
\Gamma_{5,\alpha\beta}(Q) = -\frac{1}{2}\gamma_5\bigg[
\delta_{\alpha\beta}\tilde{g}(Q^2)
-\frac{Q_\alpha Q_\beta}{4m_\Delta^2}\tilde{h}(Q^2) \bigg]\,,
\end{align}
where $\tilde{g}$ and $\tilde{h}$ are the two Poincar\'e invariant pseudoscalar form factors.

Hereafter we assume isospin symmetry and, unless otherwise noted, choose $j=3$.  The elastic weak form factors of the other $\Delta$-baryons in the multiplet can be obtained following the rules detailed in \ref{seccfcoes}.

In practice, it is useful to sum over initial- and final-state spins in order to remove the spinors in the given current:
\begin{subequations}
\label{axpscurrent}
\begin{align}
J&\,\!_{5(\mu),\lambda\omega}(K,Q)   \nonumber \\
& :=\sum_{s_f,s_i}u_\lambda(P_f;s_f){\cal J}_{5(\mu)}(K,Q)\bar{u}_\omega(P_i;s_i)\\
& =\Lambda_+(P_f)R_{\lambda\alpha}(P_f)\Gamma_{5(\mu),\alpha\beta}(Q)\Lambda_+(P_i)R_{\beta\omega}(P_i)\,,
\end{align}
\end{subequations}
where the positive-energy spinor projector and Rarita-Schwinger projection operator are, respectively:
\begin{subequations}
\begin{align}
\Lambda_+(P) & =  \frac{1}{2}\bigg( {\mathbb I}_{\rm D} + \gamma\cdot \hat P\bigg)\,,\\
\nonumber
R_{\mu\nu}(P) &  = \delta_{\mu\nu}{\mathbb I}_{\rm D}-\\
&\frac{1}{3}\gamma_\mu\gamma_\nu+\frac{2}{3}\hat{P}_\mu\hat{P}_\nu{\mathbb I}_{\rm D}-\frac{i}{3}[\hat{P}_\mu\gamma_\nu-\hat{P}_\nu\gamma_\mu]\,,
\end{align}
\end{subequations}
with $\hat{P}_\mu=P_\mu/(im_\Delta)$.  Using Eq.\,\eqref{axpscurrent}, one obtains the desired form factors by sensibly chosen matrix projection operations -- see \ref{secffproj}.


\subsection{PCAC and Goldberger-Treiman relations}
\label{secpcac}

Using Ward-Green-Takahashi identities, one can obtain the following partially conserved axial-vector current (PCAC) relation between the current operators:
\begin{align}
\label{pcacop}
\partial_\mu {\mathpzc A}^j_{5\mu}(x)+2m_q{\mathpzc P}^j_{5}(x)=0\,.
\end{align}
Evaluating the expectation value of this current,  one obtains the $\Delta$-baryon PCAC relation:
\begin{align}
\label{pcaccurrent}
Q_\mu J_{5\mu,\lambda\omega}(K,Q)+2im_q J_{5,\lambda\omega}(K,Q)=0\,,
\end{align}
which entails
\begin{align}
\label{ffpcac0}
\nonumber
&\delta_{\alpha\beta}\big(g_1-\frac{Q^2}{4m_\Delta^2} g_3\big) -
\frac{Q_\alpha Q_\beta}{4m_\Delta^2}\big(h_1-\frac{Q^2}{4m_\Delta^2} h_3\big) \\
= &\frac{m_q}{m_\Delta}\big(\delta_{\alpha\beta}\tilde{g}
-\frac{Q_\alpha Q_\beta}{4m_\Delta^2}\tilde{h}\big)\,.
\end{align}

Considering the diagonal ($\alpha=\beta$) and non-diagonal ($\alpha\neq\beta$) components of Eq.\,\eqref{ffpcac0} separately, one finds the following two independent PCAC relations at the form factor level \cite{Alexandrou:2013opa}:
\begin{subequations}
\label{ffpcac}
\begin{align}
\label{ffpcacg}
g_1 - \frac{Q^2}{4m_\Delta^2} g_3 &= \frac{m_q}{m_\Delta}\tilde{g}\,,\\
\label{ffpcach}
h_1 - \frac{Q^2}{4m_\Delta^2} h_3 &= \frac{m_q}{m_\Delta}\tilde{h}\,.
\end{align}
\end{subequations}
Notably, Eqs.\,\eqref{ffpcac} are consequences of the operator relation, Eq.\,\eqref{pcacop}.  So, only results that comply with these identities can be called realistic; and no tuning of any element in a given calculation may be employed to secure these outcomes.

Expanding on the nucleon case \cite{Eichmann:2011pv, Chen:2021guo}, one may define two $\pi$-$\Delta$ form factors, $G_{\pi\Delta\Delta}$, $H_{\pi\Delta\Delta}$:
\begin{subequations}
\label{pidddef}
\begin{align}
\label{piddgdef}
\tilde{g}(Q^2) &=: \frac{m_\pi^2}{Q^2+m_\pi^2}\frac{f_\pi}{m_q}G_{\pi\Delta\Delta}(Q^2)\,,\\
\label{piddhdef}
\tilde{h}(Q^2) &=: \frac{m_\pi^2}{Q^2+m_\pi^2}\frac{f_\pi}{m_q}H_{\pi\Delta\Delta}(Q^2)\,,
\end{align}
\end{subequations}
where $f_\pi \approx 92\,$MeV is the pion leptonic decay constant.  At the pion mass pole, $Q^2+m_\pi^2=0$, the residues of $\tilde{g}$ and $\tilde{h}$ define two $\pi$-$\Delta$ coupling constants: \begin{equation}
G_{\pi\Delta\Delta}(-m_\pi^2) =: g_{\pi\Delta\Delta}\,,
\;
H_{\pi\Delta\Delta}(-m_\pi^2) =: h_{\pi\Delta\Delta}\,.
\end{equation}

In systematic analyses of low-energy phenomena, $g_{\pi\Delta\Delta}$ and $h_{\pi\Delta\Delta}$ should relate the fields of the $\pi$ and $\Delta$ in two different ways.
Using the currents, Eqs.\,\eqref{axdef} -- \eqref{psffsdef}, the PCAC relations, Eqs.\,\eqref{ffpcac}, and analyticity of $g_3$, $h_3$ in the neighbourhood of the pion pole, one immediately obtains two Goldberger-Treiman (GT) relations for the $\Delta$-baryon:
\begin{subequations}
\label{gtr}
\begin{align}
\label{gtrg}
g_1(0) &= \frac{f_\pi}{m_\Delta}G_{\pi\Delta\Delta}(0)\,,\\
\label{gtrh}
h_1(0) &= \frac{f_\pi}{m_\Delta}H_{\pi\Delta\Delta}(0)\,.
\end{align}
\end{subequations}

It is now apparent that the four axial and two pseudoscalar form factors can be divided into two classes:
$\{g_1,g_3,\tilde{g}\}$ and $\{h_1,h_3,\tilde{h}\}$.
Each class has its own, independent PCAC and GT relations.
Comparing with the nucleon, $g_1$, $h_1$ are kindred to the nucleon axial-vector form factor $G_A$;
$g_3$, $h_3$ are analogous to the induced-pseudoscalar form factor, $G_P$;
and $\tilde{g}$, $\tilde{h}$ are akin to the pseudoscalar form factor, $G_5$.


\section{$\Delta(1232)$ Faddeev equation framework}
\label{sectheo}
\begin{figure}[t]
\centerline{%
\includegraphics[clip, height=0.14\textwidth, width=0.45\textwidth]{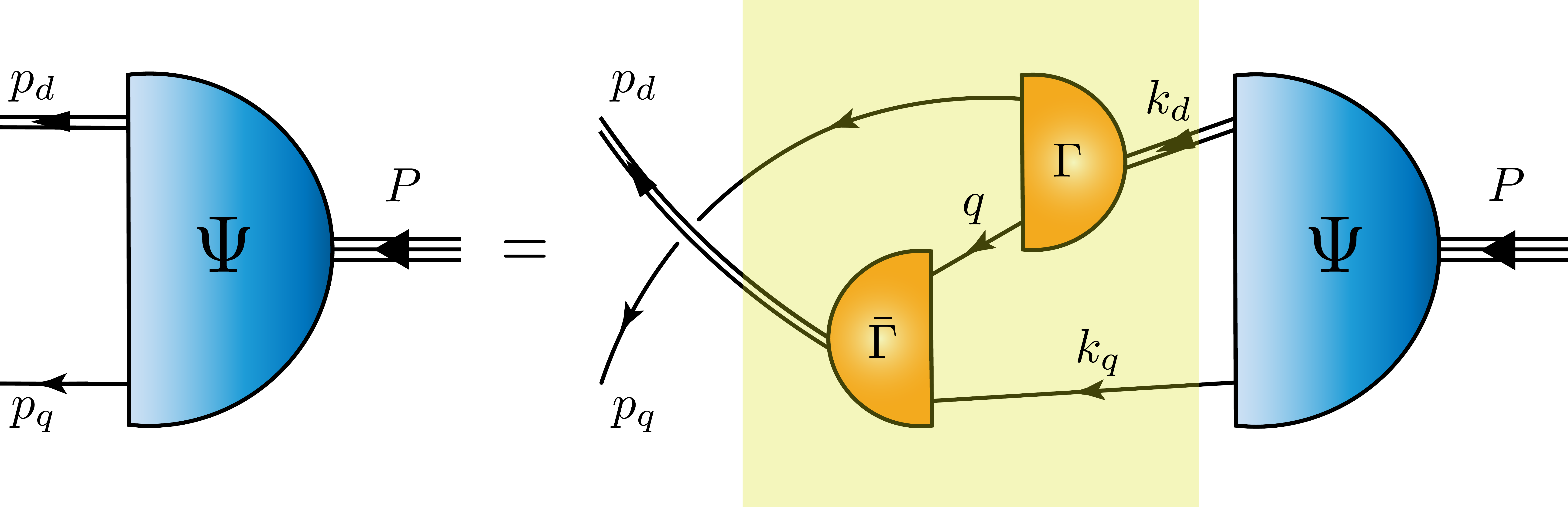}}
\caption{\label{figFaddeev}
Quark+diquark Faddeev equation.  The solution, $\Psi$, is the Poincar\'e-covariant, matrix-valued Faddeev amplitude for a baryon with total momentum $P=p_q+p_d=k_q+k_d$ constituted from three valence quarks, two of which are always contained in a nonpointlike diquark correlation.  $\Psi$ expresses the relative momentum correlation between the dressed-quarks and -diquarks.
Legend. \emph{Shaded rectangle} -- Faddeev kernel;
\emph{single line} -- dressed-quark propagator, $S$; $\Gamma$ -- diquark correlation amplitude; and \emph{double line} -- diquark propagator, ${\cal D}$.
(See \ref{secqcdmodel} for specification of these functions.)
Regarding ground-state $\Delta$-baryons, only isovector-axialvector diquarks ($\{dd\}$, $\{ud\}$, $\{uu\}$) play a material role \cite{Liu:2022ndb}.}
\end{figure}

Herein, we treat the baryon bound-state problem using a Poincar\'e-covariant quark+diquark Faddeev equation, which is sketched in Fig.\,\ref{figFaddeev}.  Crucially, the diquark correlations are nonpointlike and fully interacting; consequently, \emph{inter alia}, Fermi statistics are properly expressed.  As explained elsewhere \cite{Eichmann:2016yit, Barabanov:2020jvn}, the approach has been used widely with success.  Furthermore, to meet our goal of unifying nucleon and $\Delta$-baryon electroweak properties, we use precisely the formulation employed in Refs.\,\cite{Chen:2020wuq, Chen:2021guo, ChenChen:2022qpy}.  This ``QCD-kindred'' \linebreak approach is detailed, \emph{e.g}., in Ref.\,\cite[Appendix\,A]{Chen:2021guo}.  Nevertheless, so as to make this presentation self-contained, we reiterate some of that material in \ref{secqcdmodel}, introducing $\Delta$-baryon specific statements in place of such for the nucleon.

Regarding $I=\tfrac{3}{2}$ $\Delta$-baryons, two types of diquark correlations may be present: isovector-axialvector; and isovector-vector.  However, detailed analyses reveal \cite{Liu:2022ndb} that isovector-vector diquarks may be neglected with practically no cost.  Hence, we work with the simple isovector-axialvector Faddeev amplitude detailed in \ref{secqdqamp}.  These diquarks are characterised by the following mass-scale (in GeV):
\begin{align}
\label{axdqmass}
m_{\{uu\}_{1^+}} = m_{\{ud\}_{1^+}} = m_{\{dd\}_{1^+}} = 0.89\,,
\end{align}
whose value has been constrained by successful applications to many baryons -- see, \emph{e.g}., Refs.\,\cite{Burkert:2017djo, Chen:2018nsg, Chen:2019fzn, Lu:2019bjs, Cui:2020rmu}.

\begin{figure}[!t]
\centerline{\includegraphics[clip, height=0.33\textwidth, width=0.47\textwidth]{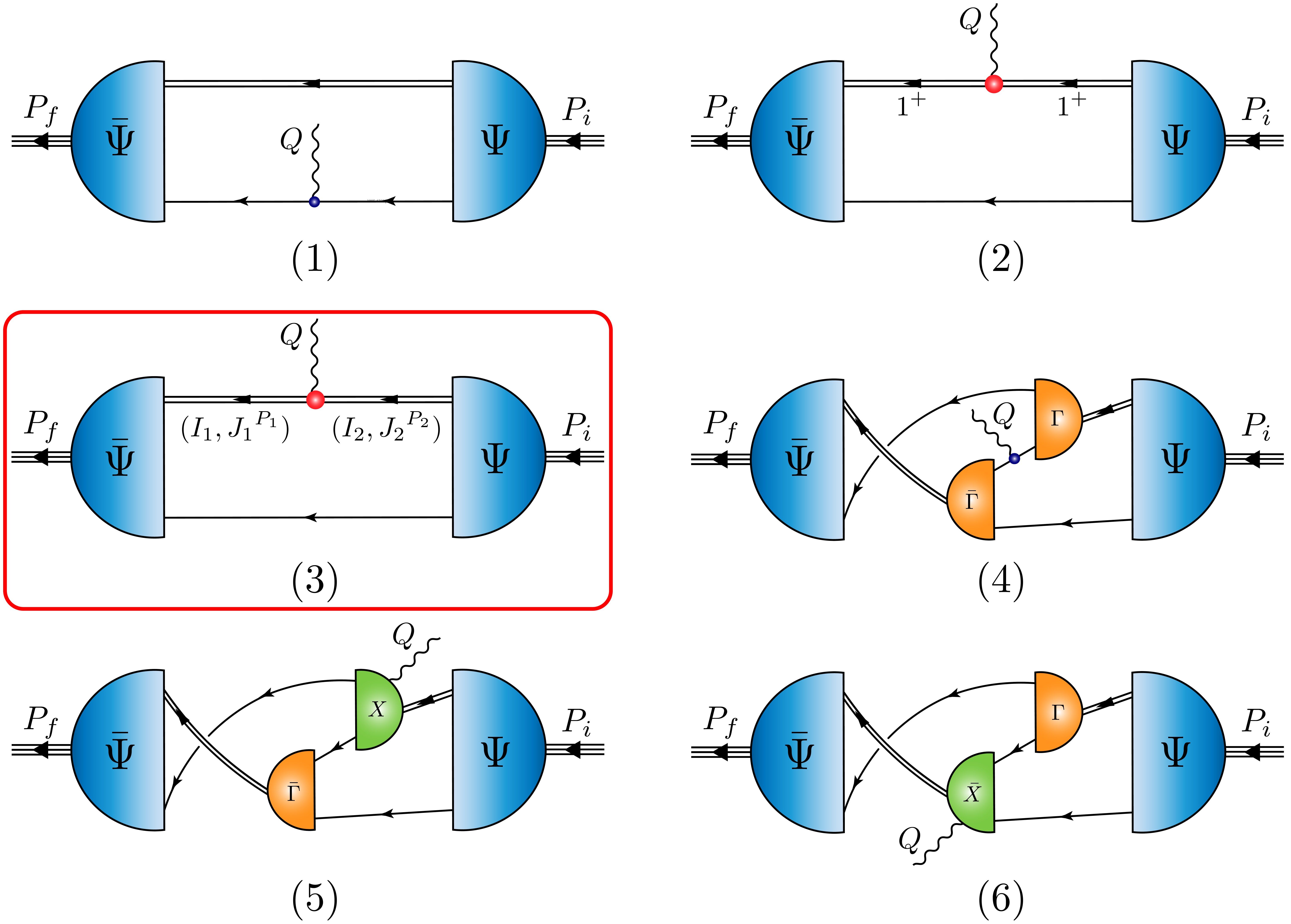}}
\caption{\label{figcurrent}
Axial or pseudoscalar current that ensures PCAC for on-shell baryons described by a Faddeev amplitude produced by the equation depicted in Fig.\,\ref{figFaddeev}.
Legend.
\emph{Single line}, dressed-quark propagator;
\emph{undulating line}, the axialvector or pseudoscalar current;
$\Gamma$, diquark correlation amplitude;
\emph{double line}, diquark propagator;
and $\chi$, seagull terms.
For $\Delta$-baryons, Diagram (3) does not contribute. }
\end{figure}

Six distinct contributions are required to provide a symmetry-preserving treatment of the axialvector and pseudoscalar currents of a baryon described by the Faddeev equation in Fig.\,\ref{figFaddeev} \cite{Chen:2020wuq, Chen:2021guo}.  For $\Delta$-baryons constituted solely from axialvector diquarks, however, Diagram (3) does not contribute because there are no other participating diquarks into which the axialvector can be transformed.  Mathematical realisations of the images in Fig.\,\ref{figcurrent} are provided in \ref{seccurr}.

\section{Results and discussion}
\label{secnum}
\subsection{Axial-vector form factors}
\label{secaxffs}
\begin{figure}[t]
\leftline{\hspace*{0.5em}{\large{\textsf{A}}}}
\vspace*{-5ex}
\hspace*{-2ex}\includegraphics[clip, width=0.48\textwidth]{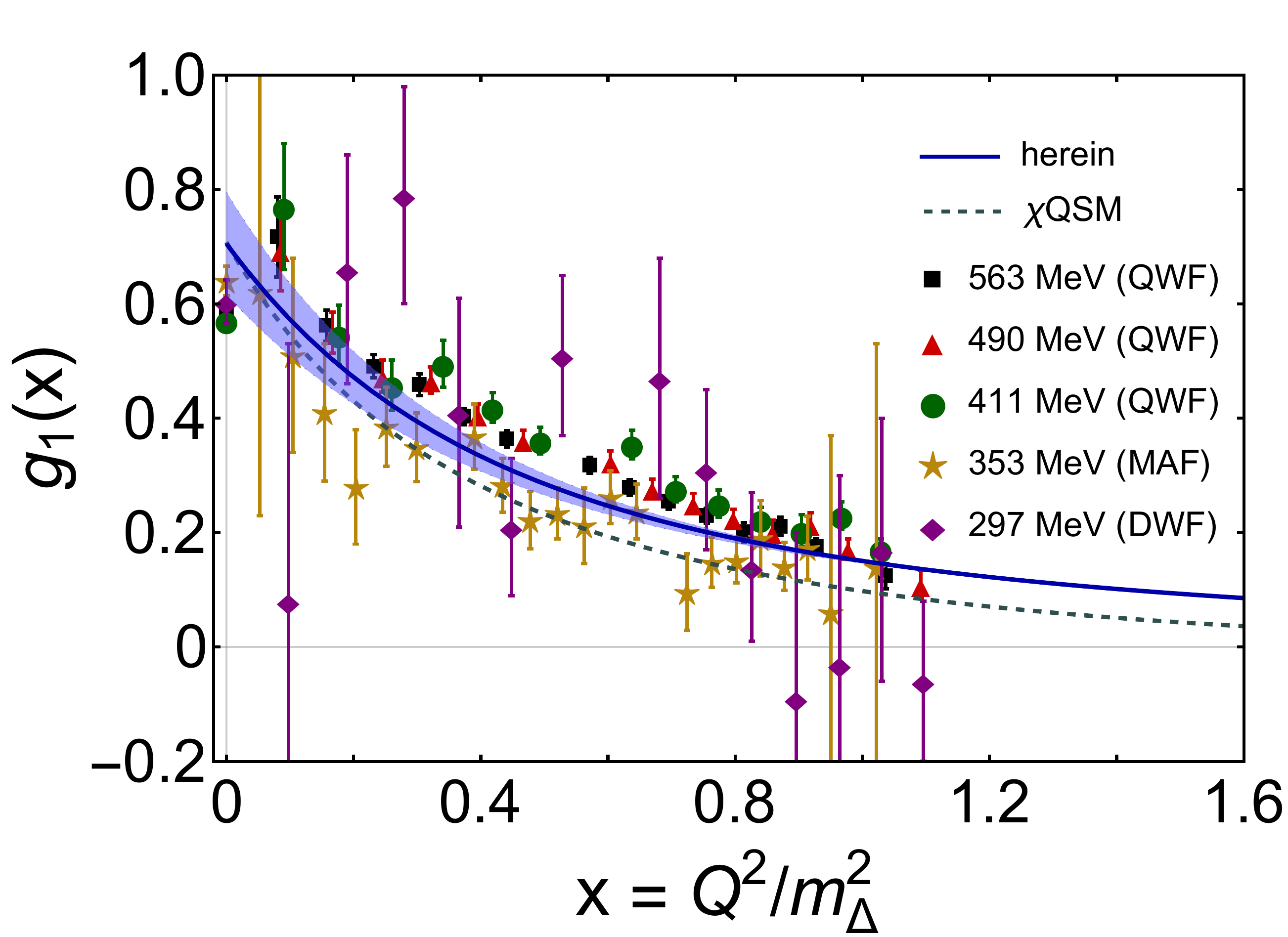}
\vspace*{1ex}
\leftline{\hspace*{0.5em}{\large{\textsf{B}}}}
\vspace*{-5ex}
\includegraphics[clip, width=0.47\textwidth]{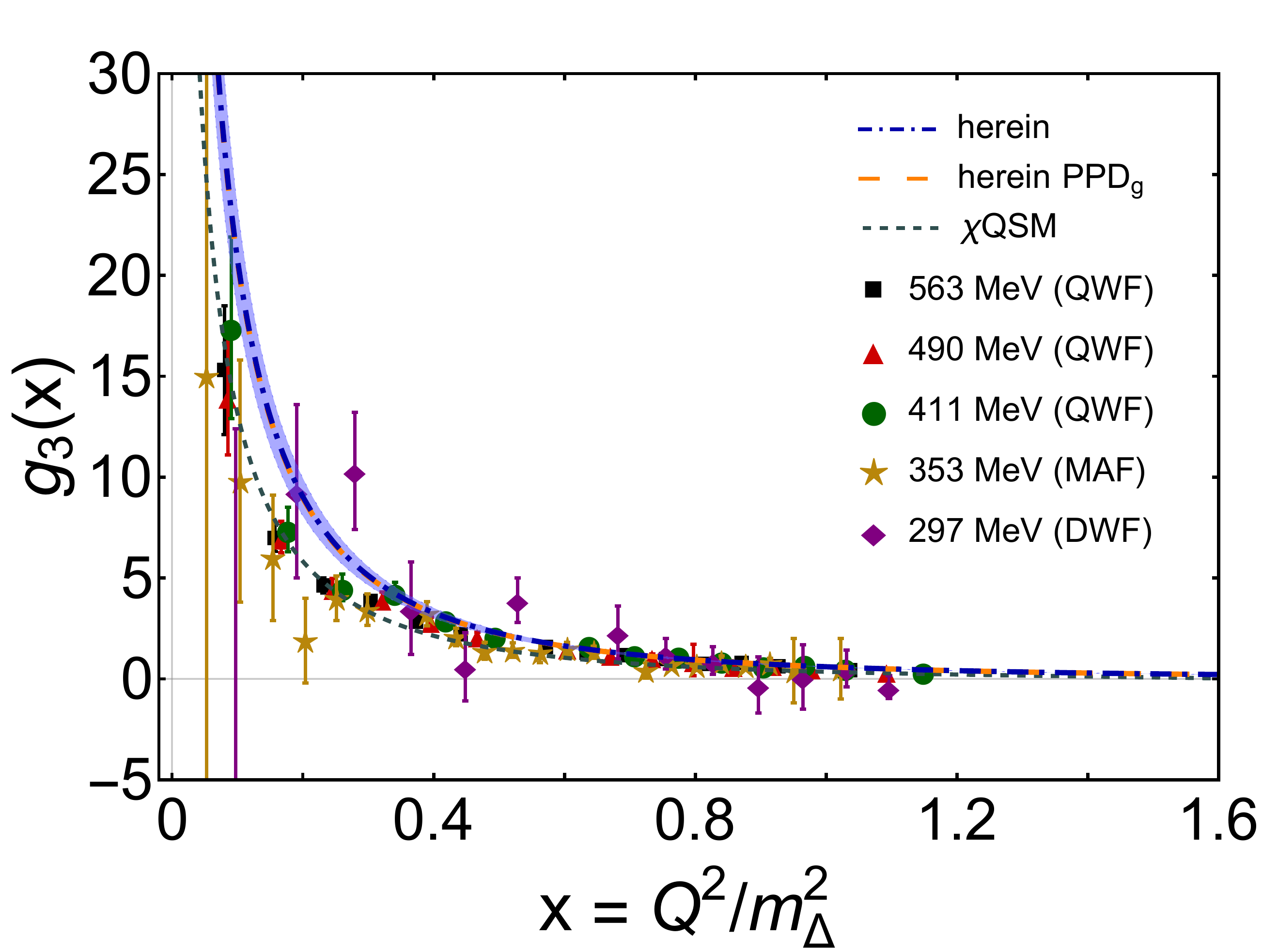}
\vspace*{4.5ex}
\caption{\label{Figg13x}
{\sf Panel A}.  $g_1(x)$ calculated herein -- blue curve within lighter blue uncertainty band.
{\sf Panel B}.  $g_3(x)$ calculated herein -- dot-dashed blue curve within lighter blue uncertainty band;
and long-dashed orange curve -- PPD approximation, Eq.\,\eqref{ppdg}, employing our CSM result for  $g_1(x)$.
Comparisons in both panels:
lQCD \cite{Alexandrou:2013opa} --
quenched (QWF) [black squares -- $m_\pi=563$\,MeV, red triangles -- $m_\pi=490$\,MeV, green circles -- $m_\pi=411$\,MeV],
mixed (MAF) [gold stars -- $m_\pi=353$\,MeV],
domain wall (DWF) [purple diamonds -- $m_\pi=297$\,MeV];
and $\chi$QSM \cite{Jun:2020lfx} -- short-dashed gray curve.
}
\end{figure}

Our predictions for $g_1(x)$, $g_3(x)$, $x=Q^2/m_\Delta^2$, are depicted in Fig.\,\ref{Figg13x}, together with results from lQCD \cite{Alexandrou:2013opa} and a $\chi$QSM \cite{Jun:2020lfx}.  As described in Sec.\,\ref{secdef}, $g_1(x)$ and $g_3(x)$ are analogues of the nucleon axial and induced pseudoscalar form factors, $G_A(x)$, $G_P(x)$, respectively.  Here and hereafter, each of our predictions is embedded in a band that expresses the impact of a $\pm 5$\% variation in the axialvector diquark mass and, consequently, the width of its correlation amplitude, Eq.\,\eqref{dqamp}.


\begin{figure}[t]
\includegraphics[clip, width=0.47\textwidth]{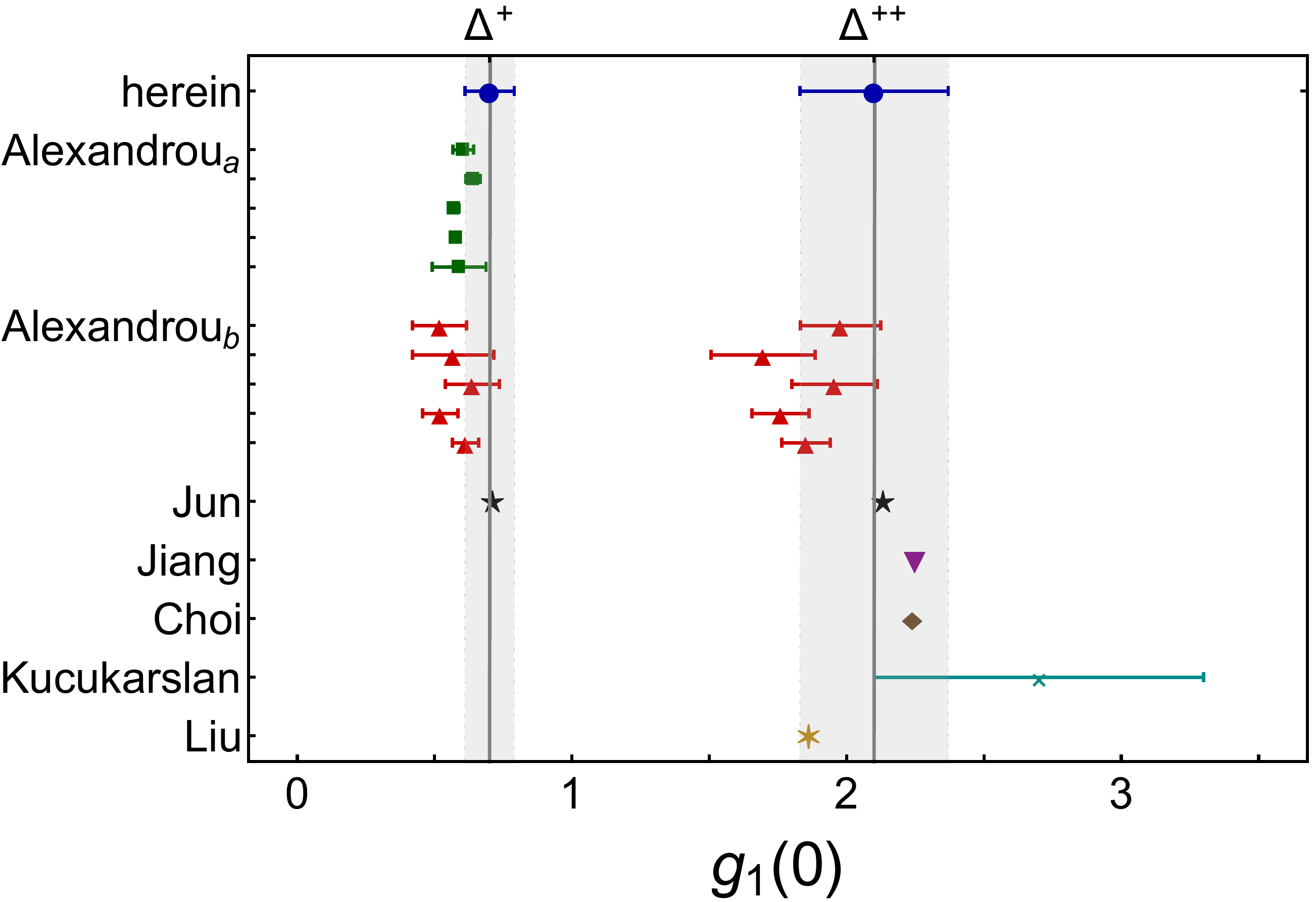}
\caption{\label{Figg10}
Our predictions for $g_1(0)$ of the $\Delta^+$ and $\Delta^{++}$ (blue circles).
They are compared with results obtained using
lQCD \cite{Alexandrou:2013opa, Alexandrou:2016xok} [green squares, red triangles, respectively -- different results correspond to different lattice setups];
a $\chi$QSM \cite{Jun:2020lfx} -- black stars;
chiral perturbation theory ($\chi$PT) \cite{Jiang:2008we} -- purple down triangle;
a relativistic constituent quark model (RCQM) \cite{Choi:2010ty} -- brown diamond;
light cone sum rules (LCSR) \cite{Kucukarslan:2014bla} -- cyan cross;
and
a perturbative chiral quark model (PTQM) \cite{Liu:2018jiu} -- gold asterisk.
}
\end{figure}

Considering Fig\,\ref{Figg13x}A,
one sees that our CSM prediction agrees qualitatively with the lQCD results.  One cannot say more because the lQCD uncertainties are too large.
Considering the $\chi$QSM result \cite{Jun:2020lfx}, which is the only other available calculation of $\Delta(1232)$-baryon axialvector form factors, there is agreement at low-$x$, but the $\chi$QSM produces a softer $x$-dependence.
Given that the CSM framework is explicitly Poincar\'e-covariant, one may reasonably expect its form factor predictions to remain reliable as $x$ increases, whereas those obtained in formulations which lack this feature are likely to degrade.

On the domain depicted, the central CSM result is accurately interpolated using Eq.\,\eqref{axpade} with the coefficients in Table\,\ref{tablepade}.  Interestingly, $g_1(x)$ can be interpolated, almost equally well, by a dipole form
\begin{align}
\label{g1dipole}
g_1(x)=\frac{g_1(0)}{\big(1+x/(m_A^\Delta/m_\Delta)^2\big)^2}\,,
\end{align}
with the axial mass $m_A^\Delta=0.95(2)m_\Delta$.
In this context, the nucleon axial mass is $m_A = 1.23(3)\,m_N$, where $m_N$ is the nucleon mass.
Evidently, converted to GeV, these dipole masses are equal within mutual uncertainties.

The $\Delta$-baryon axial charge is defined via $g_A^{\Delta}=g_1(x=0)$; and although predictions for $\Delta$-baryon axial form factors are rare, there are many calculations of $g_A^{\Delta}$, using a variety of frameworks.  In Fig.\,\ref{Figg10}, we depict our predictions:
\begin{equation}
\label{g10}
g_A^{\Delta^{+}} = 0.71(9)\,,\;
g_A^{\Delta^{++}} = 2.13(27)=3g_A^{\Delta^{+}},
\end{equation}
along with values obtained using other methods.   Evidently, there is general agreement on the results, although the lQCD values lie systematically lower than other estimates.

In Table\,\ref{tablegr}, referring to Fig.\,\ref{figcurrent}, we list the relative strengths of each diagram contribution to $g_A^{\Delta^+}$.
Diagram (1), with the weak boson striking the dressed quark in association with a spectator axialvector diquark, is dominant.
On the other hand, Diagrams (2) and (4), both contribute materially.
There is no contribution from Diagrams (5) and (6) because the seagull terms, Eqs.\,\eqref{axsg}, \eqref{axsgb}, are purely longitudinal; hence, cannot contribute to $g_1$, which is entirely determined by the $Q$-transverse part of the $\Delta$-baryon axial current -- see the $g_1$ projection, Eqs.\,\eqref{axproj} -- \eqref{axprojcoes}.

\begin{table}[!t]
\caption{\label{tablegr}
Referring to Fig.\,\ref{figcurrent}, separated diagram contributions (in \%) to $g_1(0)$, $g_3(0)$, $h_1(0)$, $h_3(0)$ and $G_{\pi\Delta\Delta}(0)$, $H_{\pi\Delta\Delta}(0)$.
Diagram (1): $\langle J \rangle^{A}_{\rm q}$ -- weak-boson strikes dressed-quark with axialvector diquark spectator.
Diagram (2): $\langle J \rangle^{AA}_{\rm qq}$ -- weak-boson strikes axialvector diquark with dressed-quark spectator.
Diagram (4): $\langle J \rangle_{\rm ex}$ -- weak-boson strikes dressed-quark ``in-flight'' between one diquark correlation and another.
Diagrams (5) and (6): $\langle J \rangle_{\rm sg}$ -- weak-boson couples inside the diquark correlation amplitude.
The listed uncertainties reflect the impact of $\pm 5$\% variations in the diquark masses in Eq.\,\eqref{axdqmass}, \emph{e.g}., $0.57_{10_\mp} \Rightarrow 0.57 \mp 0.10$.
}
\begin{center}
\begin{tabular*}
{\hsize}
{
l@{\extracolsep{0ptplus1fil}}
|l@{\extracolsep{0ptplus1fil}}
l@{\extracolsep{0ptplus1fil}}
l@{\extracolsep{0ptplus1fil}}
l@{\extracolsep{0ptplus1fil}}}\hline
 & $\langle J \rangle^{A}_{\rm q}$ &$\langle J \rangle^{AA}_{\rm qq}$ & $\langle J \rangle_{\rm ex}$ & $\langle J \rangle_{\rm sg}$ \\\hline
 $g_1(0)\ $ & $0.57_{10_\mp}$ & $0.16_{1_\pm}$ & $0.27_{9_\pm}$ & $\phantom{-}0$\\
 $g_3(0)\ $ & $0.56_{11_\mp}$ & $0.16_{1_\pm}$ & $0.39_{7_\pm}$ & $-0.10_{1_\pm}$\\
 $h_1(0)\ $ & $0.52_{7_\mp}$ & $0.26_{1_\pm}$ & $0.22_{5_\pm}$ & $\phantom{-}0$\\
 $h_3(0)\ $ & $0.55_{7_\mp}$ & $0.26_{1_\pm}$ & $0.21_{5_\pm}$ & $-0.019_{3_\pm}$\\  \hline
 $G_{\pi\Delta\Delta}(0)\ $ & $0.55_{11_\mp}$ & $0.16_{1_\pm}$ & $0.40_{8_\pm}$ & $-0.10_{2_\pm}$\\
 $H_{\pi\Delta\Delta}(0)\ $ & $0.54_{7_\mp}$ & $0.24_{1_\pm}$ & $0.24_{6_\pm}$ & $-0.020_{3_\pm}$\\  \hline
\end{tabular*}
\end{center}
\end{table}

Our prediction for $g_3(x)$ is drawn in Fig\,\ref{Figg13x}B and compared with results from lQCD and a $\chi$QSM.
%
Interpolation of our central result is obtained using Eq.\,\eqref{pspade} with the coefficients in Table\,\ref{tablepade}.
Once again, given the large lQCD uncertainties, one can only conclude that the lattice results are qualitatively consistent with our prediction.
On the other hand, in this case, one sees that the $\chi$QSM result is uniformly lower than our prediction.

Recalling now that $g_3$ is kindred to the nucleon induced pseudoscalar form factor, $G_P(x)$, one may expect a version of the pion pole dominance (PPD) approximation to be valid.  We find this to be true.  Indeed, as demonstrated by the comparison drawn in Fig\,\ref{Figg13x}B, to a good level of accuracy, one can write
\begin{align}
\label{ppdg}
g_3(x)\approx\frac{4}{x+m_\pi^2/m_\Delta^2}g_1(x)\,,
\end{align}
reproducing the form of the nucleon result \cite{Chen:2020wuq,Chen:2021guo}.
One can therefore consider Eq.\,\eqref{ppdg} to be useful as an internal consistency check on calculations of $\Delta$-baryon axial form factors.  As such, it may profitably used, \emph{e.g}., to analyse the results in Refs.\,\cite{Alexandrou:2013opa, Alexandrou:2016xok, Jun:2020lfx}.
We present a detailed discussion of the origin and applicability of Eq.\,\eqref{ppdg} in Sec.\,\ref{secppdpcac}.

\begin{figure}[t]
\leftline{\hspace*{0.5em}{\large{\textsf{A}}}}
\vspace*{-5ex}
\includegraphics[clip, width=0.47\textwidth]{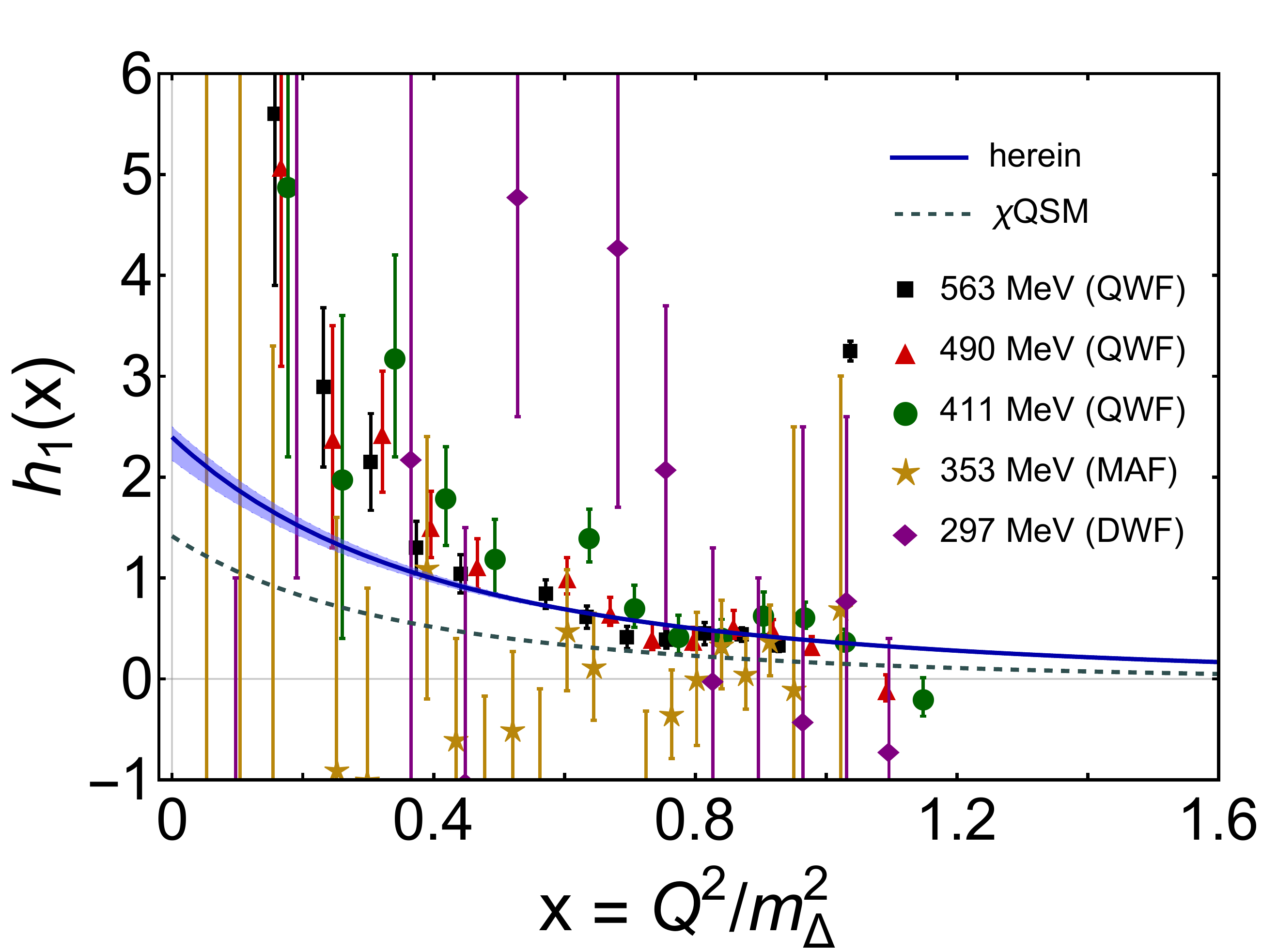}
\vspace*{1ex}
\leftline{\hspace*{0.5em}{\large{\textsf{B}}}}
\vspace*{-5ex}
\hspace*{-1ex}\includegraphics[clip, width=0.48\textwidth]{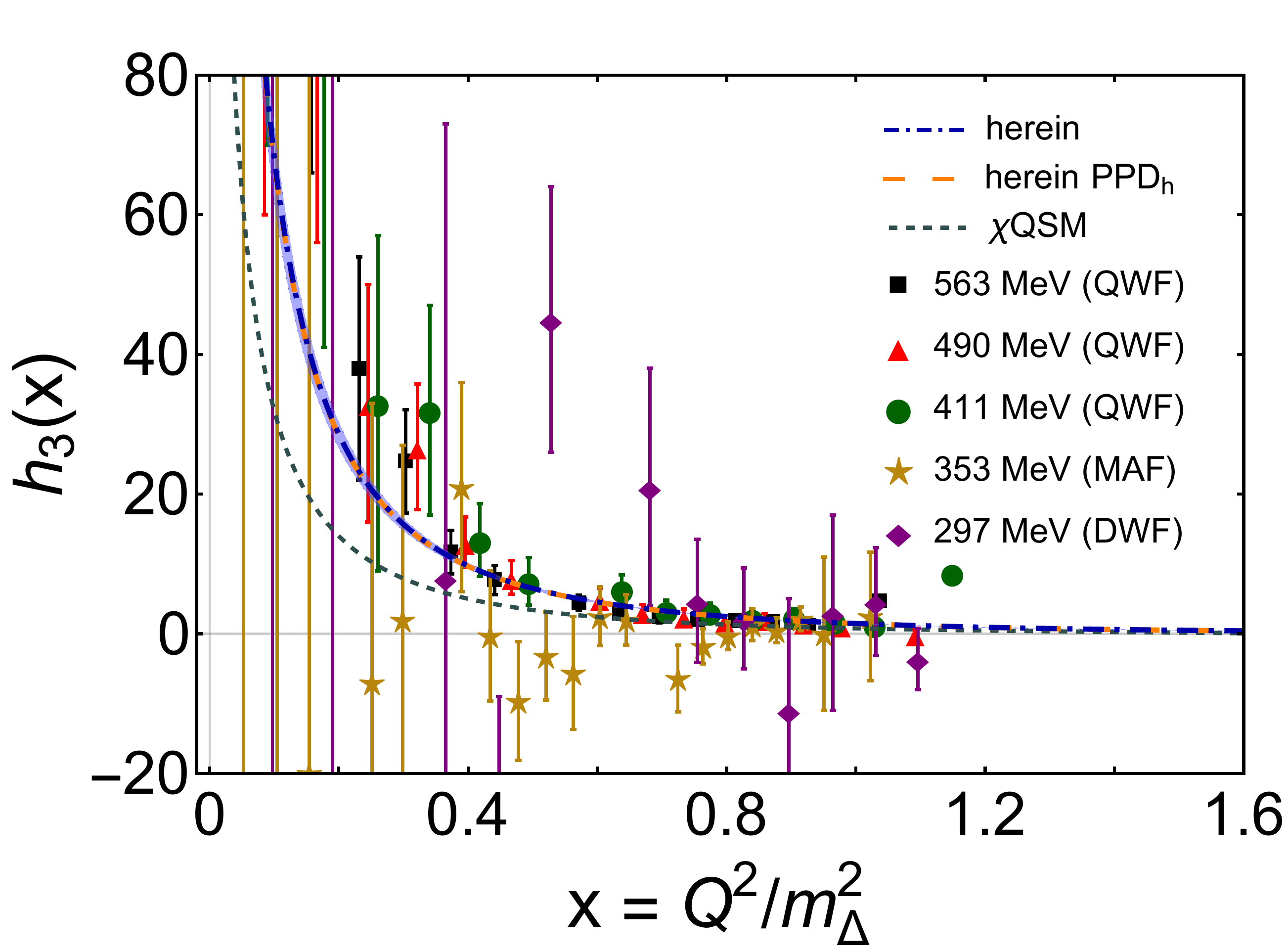}
\vspace*{4.5ex}
\caption{\label{Figh13x}
{\sf Panel A}.  $h_1(x)$ calculated herein -- blue curve within lighter uncertainty band.
{\sf Panel B}.  $h_3(x)$ calculated herein -- dot-dashed blue curve within lighter uncertainty band; and long-dashed orange curve -- PPD approximation, Eq.\,\eqref{ppdh}, employing our CSM result for $h_1(x)$.
Comparisons in both panels:
lQCD results\,\cite{Alexandrou:2013opa} -- QWF [black squares -- $m_\pi=563$\,MeV, red triangles -- $m_\pi=490$\,MeV, green circles -- $m_\pi=411$\,MeV],
MAF [golden stars -- $m_\pi=353$\,MeV],
DWF [purple diamonds -- $m_\pi=297$\,MeV];
and $\chi$QSM result\,\cite{Jun:2020lfx} -- short-dashed gray curve.
}
\end{figure}

In Row~2 of Table\,\ref{tablegr}, referring to Fig.\,\ref{figcurrent}, we list the relative strengths of each diagram contribution to $g_3(0)$.  Once again, Diagram (1), with the weak boson striking the dressed quark in association with a spectator axialvector diquark, is the dominant contributor; Diagram (2) and (4) contributions are significant; and in this case, the seagull terms act to cancel some of the Diagram (4) strength.


Our predictions for the remaining two $\Delta$-baryon axial form factors are drawn in Fig.\,\ref{Figh13x}: accurate interpolations of the central results are obtained using Eqs.\,\eqref{axpade} -- \eqref{EqR}, with the coefficients in Table\,\ref{tablepade}.

Once more, the figures compare our predictions with the only other available calculations \cite{Alexandrou:2013opa, Jun:2020lfx}.
For these two form factors, the lQCD uncertainties are especially large; so, little can be concluded from the numerical comparison.  Qualitatively, however, there are significant disagreements.
Ref.\,\cite{Alexandrou:2013opa} argues that $h_1$ should exhibit a pion simple pole and $h_3$ a pion double pole.
%
%
We disagree with these statements.
Reviewing the projection matrices, Eq.\,\eqref{axproj}, and the associated coefficients, Eq.\,\eqref{axprojcoes}, it is immediately apparent that, like $g_1$, which is regular, $h_1$ only receives contributions from ${\mathpzc s}_{3,4}$, \emph{i.e}., it is entirely determined by the $Q$-transverse part of the $\Delta$-baryon axial current; hence, cannot contain a pion pole.
Turning to $h_3$, insofar as projection matrices are concerned, this form factor is akin to $g_3$; so, must express the same pion simple pole structure.

In support of these observations we note that whilst the $\chi$QSM results are not in quantitative agreement with our predictions, their qualitative pion pole structure predictions are consistent: $h_1$ is regular and $h_3$ exhibits a simple pole.
On the domain depicted, the $\chi$QSM results for $h_{1,3}(x)$ are uniformly smaller than our predictions.  We find $h_1(0)=2.35(17)$, whereas the $\chi$QSM result is $h_1(0)=1.42$.

Like $g_{1,3}$, the characters of $h_{1,3}$ are kin to $G_{A,P}$ for the nucleon. Therefore, once again, one should anticipate a PPD relation, \emph{viz}.\
\begin{align}
\label{ppdh}
h_3(x)\approx\frac{4}{x+m_\pi^2/m_\Delta^2}h_1(x)\,.
\end{align}
In Fig\,\ref{Figh13x}B, this formula is clearly shown to provide a good approximation.
A detailed discussion of the origin and applicability of Eq.\,\eqref{ppdh} is presented in Sec.\,\ref{secppdpcac}.

In the third and fourth rows of Table\,\ref{tablegr}, we list the relative strengths of each current diagram contribution to $h_{1,3}(0)$.
There are gross similarities with the $g_{1,3}(0)$ pattern.  The differences are a reversal in the strengths of Diagrams (2) and (4) and a much smaller (in magnitude) seagull contribution to $h_3(0)$ as compared with that to $g_3(0)$.

\subsection{$\pi$-$\Delta$ form factors and GT relations}
\label{secpsffs}
Consider now the $\Delta$-baryon pseudoscalar current, $J_{5,\lambda\omega}$.
We focus on $G_{\pi\Delta\Delta}(x)$, $H_{\pi\Delta\Delta}(x)$ instead of $\tilde{g}(x)$, $\tilde{h}(x)$ because (\emph{i}) this largely eliminates sensitivity to pion mass in the results and (\emph{ii}) the former functions are renormalisation point invariant, unlike the latter two.

\begin{figure}[t]
\leftline{\hspace*{0.5em}{\large{\textsf{A}}}}
\vspace*{-5ex}
\includegraphics[clip, width=0.47\textwidth]{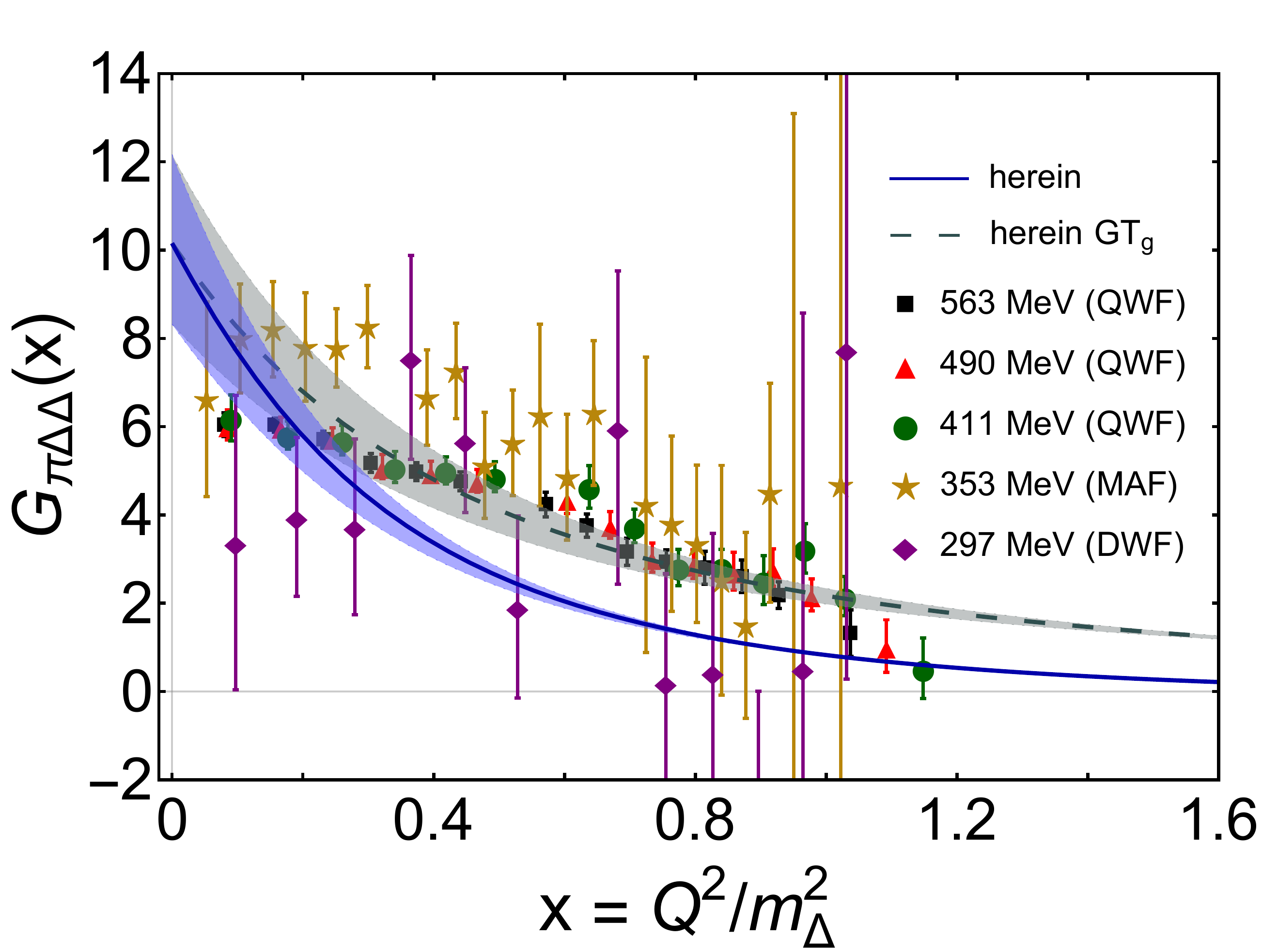}
\vspace*{1ex}
\leftline{\hspace*{0.5em}{\large{\textsf{B}}}}
\vspace*{-5ex}
\includegraphics[clip, width=0.47\textwidth]{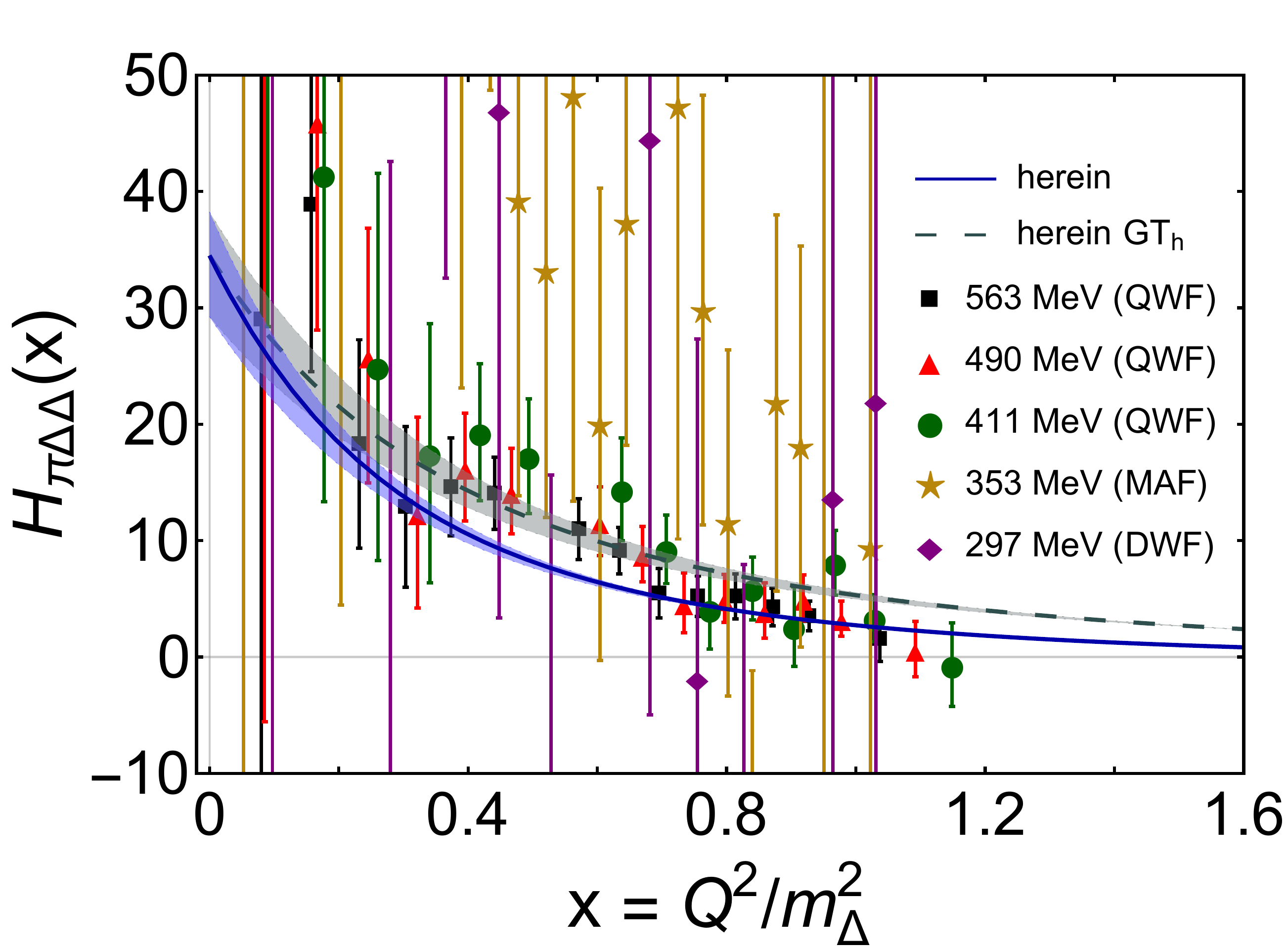}
\vspace*{4.5ex}
\caption{\label{Figps}
{\sf Panel A}.  $G_{\pi\Delta\Delta}(x)$ calculated herein -- blue curve within lighter uncertainty band;
and long-dashed grey curve within lighter grey band -- Eq.\,\eqref{gtrxgtrxg}, using CSM prediction for $g_1(x)$.
{\sf Panel B}.  $H_{\pi\Delta\Delta}(x)$ calculated herein -- dot-dashed blue curve within lighter uncertainty band.
and long-dashed grey curve within lighter grey uncertainty band -- Eq.\,\eqref{gtrxgtrxg}, using CSM prediction for  $h_1(x)$.
Comparison, both panels:
lQCD results \cite{Alexandrou:2013opa} -- QWF [black squares -- $m_\pi=563$\,MeV, red triangles -- $m_\pi=490$\,MeV, green circles -- $m_\pi=411$\,MeV], MAF [golden stars -- $m_\pi=353$\,MeV], and DWF [purple diamonds -- $m_\pi=297$\,MeV].
}
\end{figure}

Our CSM prediction for $G_{\pi\Delta\Delta}(x)$ is drawn in Fig\,\ref{Figps}A: accurate interpolation of the central result is obtained using Eq.\,\eqref{axpade} with the coefficients in Table\,\ref{tablepade}.
Furthermore, on the depicted domain, a fair approximation to the result may also be obtained with a dipole function characterised by the mass scale $\Lambda_{\pi\Delta\Delta} = 0.68(4)\,m_\Delta$ $= 0.84(5)\,$GeV, \emph{i.e}., a soft form factor.
The analogous scale for the nucleon is $0.79\,$GeV \cite{Chen:2021guo}; and just as with that analysis, our prediction for $\Lambda_{\pi\Delta\Delta}$  is $\sim 20$\% larger than, hence qualitatively equivalent to, the $\pi \Delta\Delta$ dipole mass inferred from a dynamical coupled-channels (DCC) analysis of $\pi N$, $\gamma N$ interactions \cite{Kamano:2013iva}.  This confirms that future such DCC studies may profit by implementing couplings and range parameters determined in analyses like ours.

It is worth stressing that the CSM result for $G_{\pi\Delta\Delta}(x)$ does not exhibit a pion pole contribution and, within their larger uncertainties, the lQCD results agree with this prediction.  Notwithstanding those large uncertainties, one may reasonably conclude that the CSM result is softer than that obtained using lattice regularisation.

The CSM prediction for $H_{\pi\Delta\Delta}(x)$ is depicted in Fig\,\ref{Figps}B: accurate interpolation of the central result is obtained using Eq.\,\eqref{axpade} with the coefficients in Table\,\ref{tablepade}.
The large lQCD uncertainties make it difficult to draw conclusions from any comparison.  It is plain, however, that the CSM prediction is a regular function and although Ref.\,\cite{Alexandrou:2013opa} argues for a pion simple pole in this function, there is little signal of this in the lattice results.

Any sensible calculation of $G_{\pi\Delta\Delta}(x)$ and $H_{\pi\Delta\Delta}(x)$ should satisfy the GT relations, Eqs.\,\eqref{gtr}.  Checking this, we obtain
\begin{subequations}
\label{gtrnum}
\begin{align}
G_{\pi\Delta\Delta}(0) & = 10.16(1.83) \nonumber \\
& {\rm cf.}\; \frac{m_\Delta}{f_\pi}g_1(0) = 10.42(1.32)\,,\\
H_{\pi\Delta\Delta}(0) &=  34.50(3.74) \nonumber \\
& {\rm cf.}\; \frac{m_\Delta}{f_\pi}h_1(0) = 34.48(2.49)\,;
\end{align}
\end{subequations}
so, our results comply with the GT constraints.

Extrapolating $G_{\pi\Delta\Delta}$ and $H_{\pi\Delta\Delta}$ to $Q^2=-m_\pi^2$, we find the two distinct $\pi$-$\Delta$ couplings
\begin{subequations}
\label{piddcpl}
\begin{align}
\label{gpiddva}
g_{\pi\Delta\Delta} &:= G_{\pi\Delta\Delta}(Q^2=-m_\pi^2) = 10.46(1.88)\,,\\
h_{\pi\Delta\Delta} &:= H_{\pi\Delta\Delta}(Q^2=-m_\pi^2) = 35.73(3.75)\,.
\end{align}
\end{subequations}
Regarding $g_{\pi\Delta\Delta}$, a forty-year-old near-threshold $\pi^-p\to\pi^+\pi^-n$  experiment places only a very loose constraint \cite{Arndt:1979cq}: $1.1\lesssim g_{\pi \Delta\Delta}\lesssim 30$.
One may also compare with model calculations:
$g_{\pi\Delta\Delta} \approx 14.3$ \cite[quark model -- Eq.\,(B.21)]{Brown:1975di};
$ (9/5)g_{\pi NN} = 23.7(6)$ \cite[baryon\,$1/N_c$]{Dashen:1993jt}; \linebreak
$11.8(2.0)$ \cite[LCSR]{Zhu:2000zd};
$12.0$ \cite[current parametrisation]{Buchmann:2013fxa};
$12.0$ \cite[AdS/QCD model]{Wang:2015osq};
$15.88(6.04)(5.12)$ \cite[$\chi$PT]{Yao:2016vbz}.
An error-weighted average of these results is
$g_{\pi\Delta\Delta}^{\rm ewa}$ \linebreak $=$ $12.5 (1.6)$,
with which our prediction is well aligned.  (For results with no or an unrealistic error, we introduced an uncertainty equal to the relative error in the mean of the central values $=40$\%.)  Including our prediction in the analysis, the result is
\begin{equation}
g_{\pi\Delta\Delta}^{\rm ewa} = 11.6(1.2)\,.
\end{equation}
For comparison, $g_{\pi NN} = 13.2(3)$ \cite[Fig.\,11a]{Chen:2021guo}.

Using Eqs.\,\eqref{gtrnum}, \eqref{piddcpl}, one can calculate two corresponding Goldberger-Treiman discrepancies:
\begin{subequations}
\begin{align}
\Delta^g_{\rm GT} &:= 1- \frac{G_{\pi\Delta\Delta}(0)}{G_{\pi\Delta\Delta}(-m_\pi^2)}=0.029(0)\,,\\
\Delta^h_{\rm GT} &:= 1- \frac{H_{\pi\Delta\Delta}(0)}{H_{\pi\Delta\Delta}(-m_\pi^2)}=0.035(3)\,.
\end{align}
\end{subequations}
These differences measure the deviation of the on-shell results for $G_{\pi\Delta\Delta}$, $H_{\pi\Delta\Delta}$ from their chiral limit values.  Evidently, these discrepancies are modest and commensurate with that predicted for the nucleon \cite{Chen:2021guo}: $0.030(1)$.

We would like to stress that symmetry only requires that the GT relations, Eqs.\,\eqref{gtr}, are satisfied on $x\simeq 0$.  To illustrate their domain of approximate utility, the panels in Fig.\,\ref{Figps} also display the following two functions:
\begin{equation}
\label{gtrxgtrxg}
G^\prime_{\pi\Delta\Delta}(x) = \frac{m_\Delta}{f_\pi}g_1(x)\,,\;
H^\prime_{\pi\Delta\Delta}(x) = \frac{m_\Delta}{f_\pi}h_1(x)\,.
\end{equation}
Where these curves overlap with our predictions for $G_{\pi\Delta\Delta}(x)$, $H_{\pi\Delta\Delta}(x)$, one has a domain of useful approximation.  That domain is small.  A somewhat different conclusion is suggested by Ref.\,\cite[Figs.\,9, 10]{Alexandrou:2013opa}, with the GT relations being satisfied (within large uncertainties) on a material $x$ domain.  However, those outcomes are likely the result of lattice artefacts.

\begin{figure}[t]
\centerline{%
\includegraphics[clip, width=0.47\textwidth]{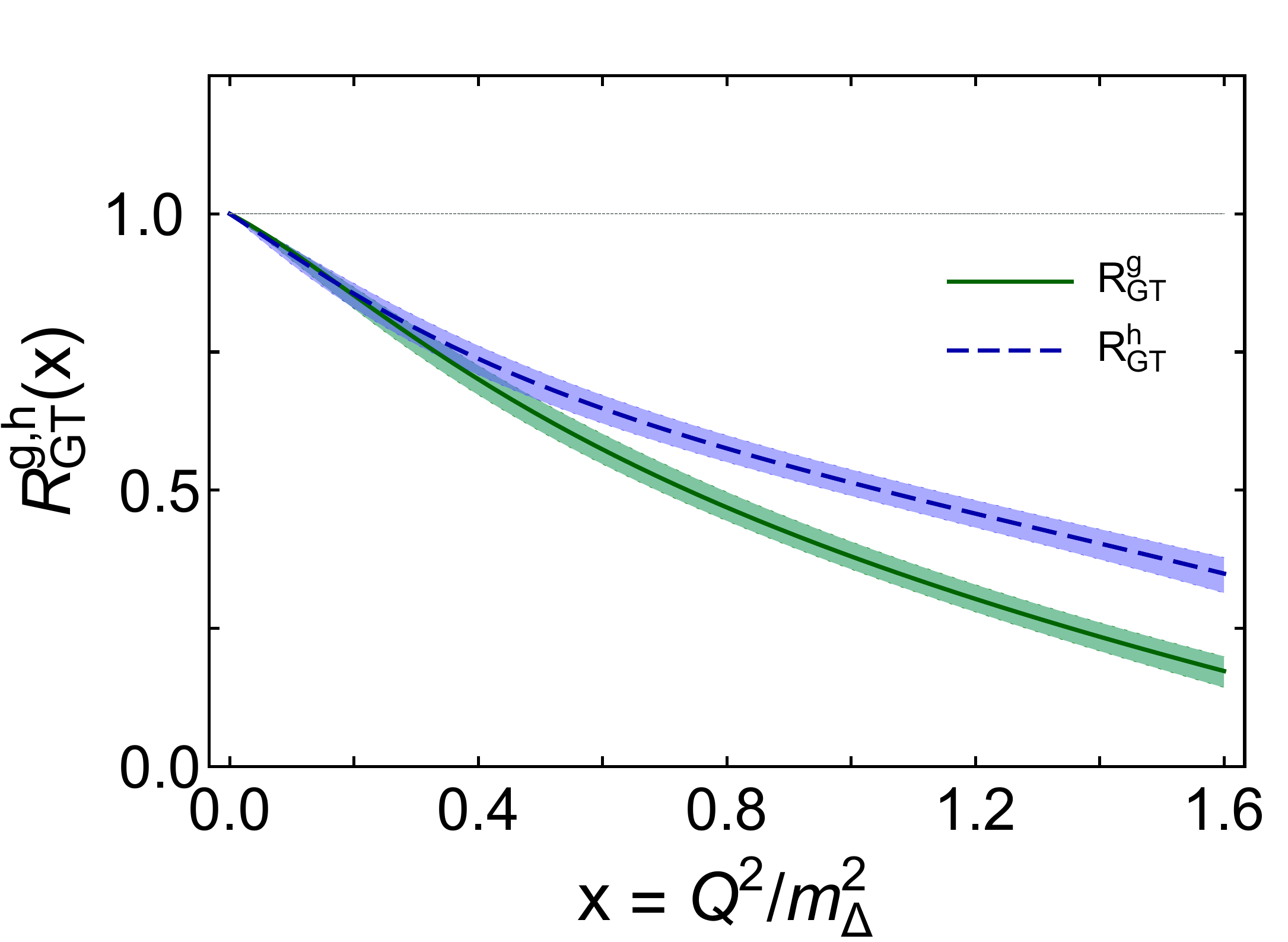}}
\caption{\label{Figrgt}
GT ratios in Eq.\,\eqref{rgtrgtg}:
$R_{\rm GT}^{g}(x)$ -- solid green curve within lighter uncertainty band; and $R_{\rm GT}^h(x)$ -- dashed blue curve within lighter uncertainty band.
}
\end{figure}

In order to explicate the domain of approximate validity, Fig.\,\ref{Figrgt} depicts the following GT ratios:
\begin{equation}
\label{rgtrgtg}
R_{\rm GT}^g(x) = \frac{f_\pi G_{\pi\Delta\Delta}(x)}{m_\Delta g_1(x)}\,,\;
R_{\rm GT}^h(x) = \frac{f_\pi H_{\pi\Delta\Delta}(x)}{m_\Delta h_1(x)}\,.
\end{equation}
These curves decreasing monotonically from unity as $x$ increases from zero, each deviating from unity by more than 10\% on $x>0.14$ and more than 20\% on $x>0.29$.

In the last two rows of Table\,\ref{tablegr}, referring to Fig.\,\ref{figcurrent}, we list the relative strengths of each diagram contribution to $G_{\pi\Delta\Delta}(0)$, $H_{\pi\Delta\Delta}(0)$, respectively.
Notably, the breakdown for $G_{\pi\Delta\Delta}(0)$ is very much like that for $g_3(0)$; and the separation for $H_{\pi\Delta\Delta}(0)$ strongly resembles the $h_3(0)$ pattern.
Similar statements were also true for the nucleon induced-pseudoscalar and true pseudoscalar form factors, $G_{P,5}$, respectively -- see Ref.\,\cite[Table~1]{Chen:2021guo}; and the explanation is the same.
Namely, if one focuses on the singular (longitudinal) part of the axial current, $J_{5\mu,\lambda\omega}$, which provides the overwhelmingly dominant contribution to $g_3(0)$, $h_3(0)$,
and compares the related projection matrices for $g_3$, $h_3$, $\tilde{g}$, $\tilde{h}$ -- see Eqs.\,\eqref{axprojffs}, \eqref{axprojcoes}, \eqref{psprojffs}, \eqref{psprojcoes},
then the following correspondences become apparent:
\begin{subequations}
\begin{align}
g_3(0) &\approx \tilde g(0) \propto G_{\pi\Delta\Delta}(0)\,, \\
h_3(0) & \approx \tilde h(0) \propto H_{\pi\Delta\Delta}(0)\,.
\end{align}
\end{subequations}
Hence, the relative strengths of different diagram contributions must be approximately the same in each case.

\subsection{Dissecting the PCAC and PPD relations}
\label{secppdpcac}
Equation~\eqref{pcacop} is an operator relation.  Thus, any physical results for axialvector and pseudoscalar form factors should satisfy Eqs.\,\eqref{ffpcac}.
In Ref.\,\cite{Chen:2021guo}, a theoretical framework was constructed which guarantees the analogous outcomes for the nucleon -- see Appendix\,D therein for a proof.
Herein, we have adapted that approach to the $\Delta$-baryon; and, using the explicit expressions for the current in Fig.\,\ref{figcurrent}, written in \ref{seccurr}, and following the same steps as for the nucleon, one may readily establish algebraically that all our results comply with Eqs.\,\eqref{ffpcac}.

\begin{figure}[t]
\centerline{%
\includegraphics[clip, width=0.47\textwidth]{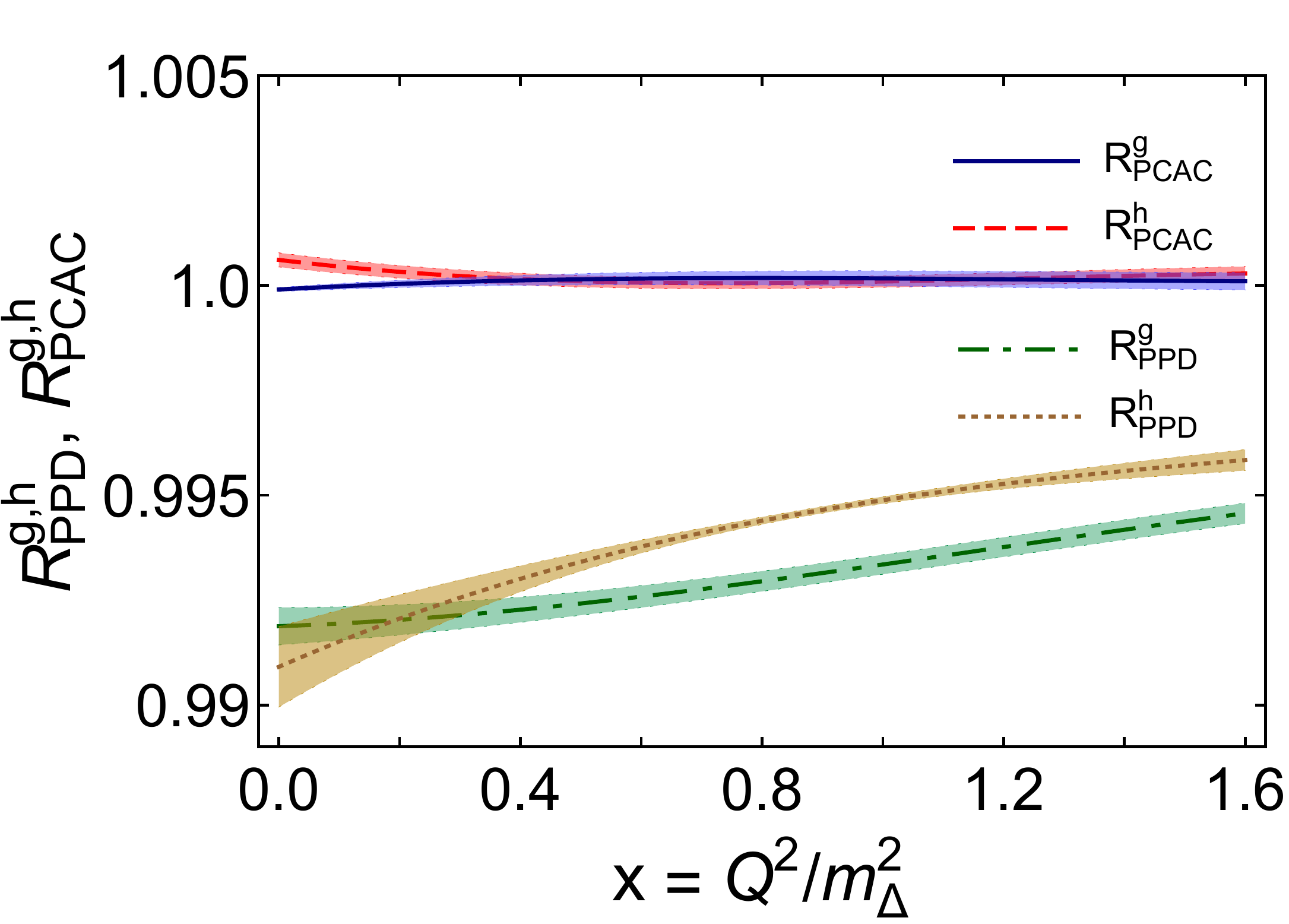}}
\caption{\label{figrpcacppd}
Numerical verification of the PCAC relations in Eqs.\,\eqref{rpcac}: solid blue curve within lighter band -- Eq.\,\eqref{rpcacg}; and dashed red curve within lighter band -- Eq.\,\eqref{rpcach}.
In addition, numerical check of the PPD relations in Eqs.\,\eqref{rppd}: dot-dashed green curve within lighter band -- Eq.\,\eqref{rppdg}; and dotted gold curve within lighter band -- Eq.\,\eqref{rppdh}.
As usual, the bands indicate the response to a $\pm 5$\% variation in the mass of the axialvector diquark.
}
\end{figure}

Notwithstanding that, numerical verification is also useful, not least because it reveals the level of accuracy in our calculations.  Therefore, consider the following two $\Delta$-baryon PCAC ratios:
\begin{subequations}
\label{rpcac}
\begin{align}
\label{rpcacg}
R_{\rm PCAC}^g & = \frac{4g_1}{ x \, g_3+4 \tilde{g} \, m_q/m_\Delta}\,,\\
\label{rpcach}
R_{\rm PCAC}^h & = \frac{4h_1}{x \, h_3+4 \tilde{h}\, m_q/m_\Delta}\,.
\end{align}
\end{subequations}
Our calculated results for both are drawn in Fig.\,\ref{figrpcacppd}: on the entire displayed domain, both curves are practically indistinguishable from unity.  We reiterate that these outcomes are parameter independent.

Two PPD relations were introduced and discussed in Sec.\,\ref{secaxffs} -- Eqs.\,\eqref{ppdg}, \eqref{ppdh}.  In order to draw additional links with nucleon properties, we reconsider them here from a different perspective.  Consider the following two ratios ($\mu^\pi_\Delta = m_\pi^2/m_\Delta^2$):
\begin{subequations}
\label{rppd}
\begin{align}
\label{rppdg}
R_{\rm PPD}^g & = \frac{4g_1}{(x + \mu^\pi_\Delta)\, g_3}\,,\\
\label{rppdh}
R_{\rm PPD}^h & = \frac{4h_1}{(x+\mu^\pi_\Delta)\, h_3}\,.
\end{align}
\end{subequations}
The calculated curves are depicted in Fig.\,\ref{figrpcacppd}.  Similar to the nucleon result \cite[Fig.\,8]{Chen:2021guo}, these curves lie $\lesssim 1$\% below unity on $x\simeq 0$ and grow toward unity as $x$ increases.  The behaviour is genuine, can readily be explained within our quark+diquark framework, and is actually universal for all baryon PPD ratios.  We will exemplify these things using $R_{\rm PPD}^g$.

First note that in the chiral limit, $R_{\rm PPD}^g$ -- Eq.\,\eqref{rppdg} and $R_{\rm PCAC}^g$ -- Eq.\,\eqref{rpcacg} are equivalent; hence, both are precisely unity:
\begin{align}
\label{rppdgunity}
R_{\rm PPD}^{g,m_q=0}=R_{\rm PCAC}^{g,m_q=0}	=\frac{4 g_1^{m_q=0}}{x g_3^{m_q=0}}=1\,.
\end{align}

Second, considering the dressed-quark vertex -- \linebreak Eq.\,\eqref{axvtx},
the diquark vertex -- Eq.\,\eqref{dqvxrev},
and the seagull terms -- Eqs.\,\eqref{axsg}, \eqref{axsgb},
one can establish that the axialvector current, $J_{5\mu,\lambda\omega}$, is a sum of well-defined regular and singular pieces, in consequence of which one may write
\begin{subequations}
\begin{align}
g_1 &= g_{1,{\rm regular}} + g_{1,{\rm singular}}\,,\\
g_3 &= g_{3,{\rm regular}} + g_{3,{\rm singular}}\,.
\end{align}
\end{subequations}

Furthermore, the regular part of $J_{5\mu,\lambda\omega}$ does not depend explicitly on the current-quark mass, $m_q$, and the singular part is proportional to $Q_\mu/(Q^2+m_\pi^2)$; hence, is purely longitudinal and does not contribute to $g_1$ -- see Eqs.\,\eqref{axproj} -- \eqref{axprojcoes}.
Consequently,
\begin{align}
\label{g1deco}
g_1 = g_{1,{\rm regular}} = g_{1,{\rm regular}}^{m_q=0} = g_{1}^{m_q=0}\,.
\end{align}
Extending these considerations,
\begin{subequations}
\label{g3deco}
\begin{align}
g_{3,{\rm regular}} &= g_{3,{\rm regular}}^{m_q=0}\,,\\
(Q^2+m_\pi^2)g_{3,{\rm singular}} &= Q^2g_{3,{\rm singular}}^{m_q=0} =: m_\Delta^2 \, {\mathpzc g}\,,
\end{align}
\end{subequations}
where ${\mathpzc g}$ is a regular function.
Inserting Eqs.\,\eqref{g1deco}, \eqref{g3deco} into Eq.\,\eqref{rppdg} and using Eq.\,\eqref{rppdgunity}, we arrive at
\begin{subequations}
\begin{align}
R_{\rm PPD}^g &= \frac{4g_1^{m_q=0}}{x g_3^{m_q=0} + g^{m_q=0}_{3,{\rm regular}} \mu^\pi_\Delta}\\
&= \frac{x g_3^{m_q=0}}{x g_3^{m_q=0} + g^{m_q=0}_{3,{\rm regular}} \mu^\pi_\Delta}\\
&= \frac{ x g_{3,{\rm regular}}^{m_q=0}+{\mathpzc g}}{ x g_{3,{\rm regular}}^{m_q=0}+{\mathpzc g}+g^{m_q=0}_{3,{\rm regular}}\mu^\pi_\Delta}\,.\end{align}
\end{subequations}
Plainly, on $x\simeq 0$, $R_{\rm PPD}^g$ must differ from unity because of the denominator term $\propto \mu^\pi_\Delta= m_\pi^2/m_\Delta^2$; and the size of the correction diminishes as $\mu^\pi_\Delta/ x$ with increasing $x$.

The size of the $x\simeq 0$ deviation is readily estimated algebraically.   Using Eqs.\,\eqref{ffpcacg}, \eqref{piddgdef}, \eqref{gtrg}, one finds
\begin{equation}
g_3(x) = \frac{4}{x}g_1(x)\bigg[1-\frac{\mu^\pi_\Delta}{x+\mu^\pi_\Delta}
\frac{G_{\pi\Delta\Delta}(x)/G_{\pi\Delta\Delta}(0)}{g_1(x)/g_1(0)}\bigg]\,.
\end{equation}
Now, referring to Eq.\,\eqref{rppdg}, define
\begin{equation}
g_3^{\pi}(x) = \frac{4}{x+\mu^\pi_\Delta}g_1(x)=
\frac{4}{x}g_1(x)\bigg[1-\frac{\mu^\pi_\Delta}{x+\mu^\pi_\Delta}\bigg]\,.
\end{equation}
Then,
\begin{subequations}
\begin{align}
& R^g_{\rm PPD}\equiv\frac{g^{\pi}_3(x)}{g_3(x)}\\
&=\bigg[1+\frac{\mu^\pi_\Delta}{x}
\bigg(1-\frac{G_{\pi\Delta\Delta}(x)/G_{\pi\Delta\Delta}(0)}{g_1(x)/g_1(0)}\bigg)\bigg]^{-1}\,.
\end{align}
\end{subequations}
Consequently,
\begin{align}
\label{ppdgde0}
R^g_{\rm PPD}(x\simeq0)=\big(1+\tfrac{1}{6}m_\pi^2[r_{G_{\pi\Delta\Delta}}^2 - r_{g_1}^2]\big)^{-1}\,,
\end{align}
where $r_{G_{\pi\Delta\Delta}}$, $r_{g_1}$  are form factor radii, defined, via  ($F\in\{G_{\pi\Delta\Delta}, g_1\}$)
\begin{align}
\label{radius}
r_F^2 = -6\frac{1}{m_\Delta^2}\frac{d}{dx} \frac{G_F(x)}{G_F(0)} \bigg|_{x=0}	\,.
\end{align}

Using our interpolations of $G_{\pi\Delta\Delta}$, $g_1$ -- \ref{secintpl}, one obtains
$r_{G_{\pi\Delta\Delta}} = 0.60(1)\,{\rm fm}$,
$r_{g_1} = 0.53(1)\,{\rm fm}$; and inserting these values into Eq.\,\eqref{ppdgde0}:
\begin{align}
R^g_{\rm PPD}(Q^2\simeq0) = 0.993(0).
\end{align}
This result matches and explains that in Fig.\,\ref{figrpcacppd}.
(It is worth noting here that $r_{g_1}=0.78(1)\,r_A^N$, \emph{i.e}., the axial radius of the $\Delta$-baryon is roughly 20\% smaller than that of the nucleon.)

The behaviour of the second PPD ratio, $R^h_{\rm PPD}$ -- Eq.\,\eqref{rppdh}, can similarly be explained.  Using Eq.\,\eqref{radius}, $r_{H_{\pi\Delta\Delta}} = 0.64(2)\,{\rm fm}$, $r_{h_1} = 0.56(3)\,{\rm fm}^{-1}$; hence,
\begin{subequations}
\begin{align}
\label{ppdhde0}
R^h_{\rm PPD}(x\simeq0) & =
\big(1+\frac{1}{6}m_\pi^2[r_{H_{\pi\Delta\Delta}}^2 - r_{h_1}^2]\big)^{-1}\\
& = 0.992(1)\,.
\end{align}
\end{subequations}
As expected, this value matches and explains the associated result in Fig.\,\ref{figrpcacppd}.



\section{Axial charge flavour separation}
\label{SecFlavourSep}
As noted in the Introduction, the Poincar\'e-covariant Faddeev wave functions of $\Delta$-baryons are simpler than that of the proton because $\Delta$ states only contain isovec\-tor-axialvector diquarks whereas the proton also contains isoscalar-scalar diquarks that can mix with isovec\-tor-axialvector diquarks under weak interactions.  Nevertheless, Poincar\'e-covariant $\Delta$-baryon wave functions are not trivial: in addition to $\mathsf S$-wave components, they contain significant $\mathsf P$-wave and $\mathsf S \otimes \mathsf P$-interference components \cite[Fig.\,8a]{Liu:2022ndb}.  Consequently, as with the proton \cite{Cheng:2023kmt}, there is no reference frame in which the total $J=\tfrac{3}{2}$ of the $\Delta$-baryon is merely the sum of three parallel $J=\tfrac{1}{2}$ quark spins.

These remarks can be quantified by presenting a flavour decomposition of $g_1^\Delta(0)$.  Consider first the $\Delta^{++}$.  In this case, only the $u$-quark contributes.  There are three $u$ quarks; so, one can write
\begin{equation}
g_A^{\Delta^{++}} =: 3 g_A^{\Delta_u} = 3 \times  0.71(9) \Rightarrow g_A^{\Delta_u} = 0.71(9)\,,
\end{equation}
where the last few steps express results of our calculations -- see Eq.\,\eqref{g10}.
(Here, we have explicitly removed the valence quark number, $n_u^{\Delta^{++}}=3$, from the charge.)
Our prediction may be compared with a lQCD estimate of this charge \cite{Alexandrou:2016xok}: $g_A^{\Delta_u} = 0.59(16)$.  They agree within mutual uncertainties.
(Recall Fig.\,\ref{Figg10}, in which lQCD results are systematically lower than other estimates.)
The isoscalar axial charge of any hadron is invariant under leading-order QCD evolution \cite{Deur:2018roz, Cheng:2023kmt}.

In any simple SU$(4)$ quark model, the result here would be $3 \times g_A^{\rm QM}$, $g_A^{\rm QM}=1$.  Two conclusions are immediately apparent:
(\emph{i}) owing to spin--flavour--relative-momentum correlations expressed in the Faddeev wave function, which break SU$(4)$ symmetry, the axial charge of each dressed quark within the $\Delta$-baryon is ``quenched'';
and
(\emph{ii}), consequently,
dressed-quarks in the $\Delta^{++}$ carry only $\approx 71$\% of the baryon's total spin.
In the proton, the result is $\approx 65$\%.
%

Turning to the $\Delta^-$, only $d$ quarks contribute.
In this case, using Eqs.\,\eqref{VariousJs}, we find (in the isospin symmetry limit)
\begin{equation}
g_A^{\Delta^{-}} = - g_A^{\Delta^{++}}  =: - 3 g_A^{\Delta_d} \Rightarrow g_A^{\Delta_d} =  0.71(9)\,.
\end{equation}

Evidently,
\begin{equation}
\label{pcfDelta}
\frac{g_A^{\Delta_d}}{g_A^{\Delta_u}} = 1\; {\rm cf.} \; \frac{g_A^{p_d}}{g_A^{p_u}} = -0.64(4)\,,
\end{equation}
where the last equality expresses the result for the analogous ratio in the proton \cite{ChenChen:2022qpy}.
(Recall, herein we have removed the in-hadron valence quark number from the charge: $n_u^p = 2$, $n_d^p=1$.)
The change in sign and relative magnitudes revealed by Eq.\,\eqref{pcfDelta} highlight the impacts of the additional correlations within the proton wave function on the effective axial charges of its dressed quarks.

\section{Summary and perspective}
\label{secsum}
Using a Poincar\'e-covariant quark+diquark Faddeev \linebreak equation treatment of $\Delta$-baryons and weak interaction currents that guarantee consistency with relevant Ward-Green-Takahashi identities, we delivered the first continuum predictions for all six $\Delta$-baryon elastic weak form factors.  In doing so, we unified them with the three analogous nucleon form factors, treated using the same framework elsewhere \cite{Chen:2021guo}.  Concerning $\Delta$-baryons, there are two distinct classes of partial conservation of axial current (PCAC) and related Goldberger-Treiman (GT) relations, involving form factor sets $(g_1, g_3, G_{\pi\Delta\Delta})$, \linebreak $(h_1, h_3, H_{\pi\Delta\Delta})$, and we provided a detailed discussion of their realisations within our framework.

The $\Delta$-baryon $g_1$ axial form factor is analogous to the nucleon $G_A$ form factor.  Our calculations show that it can reliably be approximated by a dipole function on $0< Q^2 \lesssim 1.6 m_\Delta^2$, where $m_\Delta$ is the $\Delta$-baryon mass, normalised by an axial charge, which takes the value $g_A^{\Delta^+}=0.71(9)$ [Eqs.\,\eqref{g1dipole}, \eqref{g10}].  The dipole mass, $m_A^\Delta=0.95(2)m_\Delta$, is a little larger than that found in analysing $G_A$.
Our prediction for $g_1(Q^2)$ is consistent with available results from lattice-regularised QCD (lQCD) \cite{Alexandrou:2011py, Alexandrou:2013opa} [Fig.\,\ref{Figg13x}A].  It is also more precise and, therefore, given the accuracy of the kindred prediction for $G_A(Q^2)$, quite likely more reliable.

Regarding the $g_3$ form factor, which is an analogue of the nucleon induced-pseudoscalar form factor, $G_P$, we showed that it possesses a first-order pion pole.  Further in this connection, to a good level of accuracy, $g_3$ and $g_1$ are related by a pion pole dominance (PPD) approximation [Eq.\,\eqref{ppdg}, Fig.\,\ref{figrpcacppd}]; again, just as one finds for the kindred nucleon form factors.

Turning to the other class of form factors, we predicted that $h_1$ is a regular function and, like $g_3$ and $G_P$, $h_3$ exhibits a first-order pion pole [Fig.\,\ref{Figh13x}].  In these statements, which are supported by algebraic analyses, we differ with those inferred from lQCD \cite{Alexandrou:2011py, Alexandrou:2013opa}.  Notably, the only other calculation of these form factors supports our findings \cite{Jun:2020lfx}.  Unsurprisingly, given the symmetry preserving character of our analysis, to a good level of accuracy, a PPD approximation links $h_1$ and $h_3$ [Eq.\,\eqref{ppdh}, Fig.\,\ref{figrpcacppd}].

The $\Delta$-baryon pseudoscalar currents are best characterised in terms of renormalisation point invariant $\pi\Delta\Delta$ form factors: $G_{\pi\Delta\Delta}$, $H_{\pi\Delta\Delta}$ [Sec.\,\ref{secpsffs}].  We find, algebraically and numerically, that both are regular functions, just as is $G_{\pi NN}$.  These results challenge the lQCD claim that $H_{\pi\Delta\Delta}$ has a pion simple pole \cite{Alexandrou:2013opa}.  Regarding $G_{\pi\Delta\Delta}$, there are many estimates of the $Q^2+m_\pi^2=0$ (pion on-shell) value.  We predict $g_{\pi\Delta\Delta}=10.46(1.88)$, which compares favourably with an error weighted average of model estimates, \emph{viz}.\ $12.5(1.6)$.  The on-shell value of the second $\pi\Delta\Delta$ form factor is $h_{\pi\Delta\Delta}= 35.73(3.75)$.  Our results are consistent with the GT symmetry constraints -- algebraically and numerically [Eqs.\,\eqref{gtrnum}]: of course, these constraints only apply on $Q^2/m_\Delta^2 \simeq 0$.

Partly as a check on our numerical methods, we verified that the PCAC relations -- expressing key symmetries of Nature and proved algebraically within our framework, are also satisfied numerically in our calculations: the mismatch is never more than $0.1$\% [Fig.\,\ref{figrpcacppd}].  We also showed that the two $\Delta$-baryon PPD approximations are satisfied at better than $1$\% on $Q^2>0$, explaining that the $Q^2/m_\Delta^2\simeq 0$ discrepancy is real and natural [Sec.\,\ref{secppdpcac}].

Having established the hardiness of our framework, we completed a flavour decomposition of the $\Delta$-baryon axial charges [Sec.\,\ref{SecFlavourSep}].  Owing to the simplicity of $\Delta$-baryon Poincar\'e-covariant wave functions when compared to that of the proton, this was relatively straightforward.  The analysis predicts that, at the hadron scale, the dressed-quarks carry $71$\% of the $\Delta$-baryon spin, with the remainder stored in quark+diquark orbital angular momentum.  In the proton, the analogous fraction is $\approx 65$\%.  Notably, too, the additional correlations within the proton wave function produce different quenchings of the $u$ and $d$ quark axial charges.

As stated at the outset, now, with reliable predictive tool established, the natural next step is to calculate the form factors that characterise weak-interaction induced $N\to \Delta(1232)$ transitions.  Reliable predictions for these transition form factors are important in order to understand modern neutrino-nucleus scattering experiments that seek physics beyond the Standard Model.  Consequently, many estimates exist.  However, none may claim to deliver a fully Poincar\'e-covariant treatment of the process, which, simultaneously, unifies it with a large array of electroweak properties of the nucleon and $\Delta$-baryons themselves.

A longer term goal is elimination of the quark+di\-quark approximation to the Faddeev kernel, replacing the resulting Faddeev amplitude with the solution of a truly three-body equation.  Following Refs.\,\cite{Eichmann:2009qa, Wang:2018kto}, this is achievable.  However, it must also be realistic; and that challenge may require an approach which goes beyond the leading-order continuum Schwinger function method truncation of the baryon three-body problem.


\medskip
\noindent\emph{Acknowledgments}.
We are grateful to Y.-S. Jun and H.-C. Kim for providing us with the $\chi$QSM results in Ref.\,\cite{Jun:2020lfx} and for constructive comments from Z.-F.~Cui, V.\,I.~Mokeev and D.-L.~Yao.
Work supported by:
National Natural Science Foundation of China (grant nos.\ 12135007 and 12247103);
%
%
Nanjing University of Posts and Telecommunications Science Foundation (grant no.\ NY221100);
and
Deutsche Forschungsgemeinschaft \linebreak (DFG) (grant no.\ FI 970/11-1).


\appendix
\setcounter{equation}{0}

\section{Colour and flavour coefficients}
\label{seccfcoes}
The explicit form of the $\Delta$-baryon Faddeev equation pictured in Fig.\,\ref{figFaddeev} is
\begin{align}
\label{deltafaddeev}
\Psi^\Delta_{\mu\nu}& (p;P) \nonumber \\
 & =\int_{dk}{\mathpzc K}^\Delta_{\mu\lambda}(p,k,P)S(\tilde{k}_q)
 {\cal D}^{1^+}_{\lambda\sigma}(\tilde{k}_d)\Psi^\Delta_{\sigma\nu}(k;P)\,,
\end{align}
where $\int_{dk}:=\int d^kp/(2\pi)^4$;
and the Faddeev equation quark-exchange kernel is
\begin{align}
{\mathpzc K}^\Delta_{\mu\lambda}(p,k,P) = \Gamma^{1^+}_\lambda(k_r)S^{\rm T}(q)\bar{\Gamma}^{1^+}_\mu(p_r)\,,	
\end{align}
with momenta ($\eta=1/3$, $\hat \eta =1-\eta$)
\begin{equation}
\begin{array}{ll}
\tilde{p}_q = p+\eta\,P\,, &
\tilde{k}_q = k+\eta\,P\,,\\
\tilde{p}_d = -p + \hat{\eta}\,P\,, &
\tilde{k}_d = -k + \hat{\eta}\,P\,,\\
q = \tilde{p}_d - \tilde{k}_q\,, & \\
p_r = \displaystyle \frac{\tilde{k}_q-q}{2}\,, &
k_r = \displaystyle \frac{\tilde{p}_q-q}{2}\,.
\end{array}
\end{equation}

Taking the product of the flavour and colour matrices in Eq.\,\eqref{deltafaddeev}, which are given in Eqs.\,\eqref{dqamp}, \eqref{qdqamp}, and subsequently projecting onto the isospinors of the specified $\Delta$ state:
\begin{subequations}
\label{deltaunits}
\begin{align}
& {\mathpzc e}_{\Delta^{++}} = \left(\begin{array}{c} 1 \\ 0 \\ 0 \\ 0\end{array}\right)\,,
& {\mathpzc e}_{\Delta^{+}} = \left(\begin{array}{c} 0 \\ 1 \\ 0 \\ 0\end{array}\right)\,,\\
& {\mathpzc e}_{\Delta^{0\phantom{+}}} = \left(\begin{array}{c} 0 \\ 0 \\ 1 \\ 0\end{array}\right)\,,
& {\mathpzc e}_{\Delta^{-}} = \left(\begin{array}{c} 0 \\ 0 \\ 0 \\ 1\end{array}\right)\,,
\end{align}
\end{subequations}
one finds that the colour-flavour coefficient of $\Delta$-baryon Faddeev equation, Eq.\,\eqref{deltafaddeev}, is ``$-1$''.

For the form factor diagrams of Fig.\,\ref{figcurrent}, each of the flavour coefficients must be calculated separately.  One has
\begin{align}
\label{cfdia1}
\sum_{k,l=1}^3\bigg[\delta^{kl}(s_f^k)^\dagger\big(\frac{\tau^j}{2}\big)(s_f^l)\bigg]\,,
\end{align}
for the probe-quark diagram -- Diagram (1), \linebreak where $\{s_f^k|k=1,2,3\}$ are given in Eq.\,\eqref{qdqfm};
\begin{align}
\label{cfdia2}
\sum_{k,l=1}^3\bigg[
(s^k_f)^\dagger\,
(s^l_f)\,
{\rm tr}\big[(t^k_f)^\dagger\,(t^l_f)\,\big(\frac{\tau^j}{2}\big)^\dagger\big]
\bigg]\,,
\end{align}
for the probe-diquark diagram -- Diagram (2), where the diquark flavour matrices are given in Eq.\,\eqref{dqfm};
and
\begin{align}
\label{cfdia4}
\sum_{k,l=1}^3\bigg[
(s^k_f)^\dagger\,
(t^l_f)\,\big(\frac{\tau^j}{2}\big)^\dagger\,
(t^k_f)^\dagger\,
(s^l_f)\bigg]\,,
\end{align}
for the exchange diagram -- Diagram (4).

The seagull case is somewhat more complicated because one needs to treat the bystander and exchange quark legs separately.  Considering Diagram (5), the exchange leg is
\begin{align}
\label{cfdia5ex}
\sum_{k,l=1}^3\bigg[
(s^k_f)^\dagger \,(t^l_f)\,
\big(\frac{\tau^j}{2}\big)^\dagger\,(t^k_f)^\dagger
\,(s^l_f)\bigg]
\end{align}
and the bystander leg is
\begin{align}
\label{cfdia5by}
\sum_{k,l=1}^3\bigg[
(s^k_f)^\dagger \,\big(\frac{\tau^j}{2}\big)\,
(t^l_f)\,(t^k_f)^\dagger
\,(s^l_f)\bigg]\,.
\end{align}

For the conjugation, Diagram (6), the exchange leg is
\begin{align}
\label{cfdia6ex}
\sum_{k,l=1}^3\bigg[
(s^k_f)^\dagger \,(t^l_f)\,
\big(\frac{\tau^j}{2}\big)^\dagger\,(t^k_f)^\dagger
\,(s^l_f)\bigg]
\end{align}
and the bystander,
\begin{align}
\label{cfdia6by}
\sum_{k,l=1}^3\bigg[
(s^k_f)^\dagger \,
(t^l_f)\,(t^k_f)^\dagger
\,\big(\frac{\tau^j}{2}\big)\,(s^l_f)\bigg]\,.
\end{align}

Finally, again, one must project these matrices, \linebreak Eqs.\,\eqref{cfdia1} -- \eqref{cfdia6by}, into the required $\Delta$-baryon charge state using the isospin vectors in Eqs.\,\eqref{deltaunits}.

The colour factors are ``1'' for impulse-approximation contributions -- Diagrams (1) and (2), and ``$-1$'' for the exchange and seagull diagrams -- Diagrams (4), (5), (6).

Using the $\Delta^+$-baryon as the exemplar, writing the flavour and colour factors explicitly, one has
\begin{align}
\nonumber
J^{\Delta^+}_{5(\mu),\lambda\omega}=\frac{1}{3}\big(&J^{{\rm q}(1)}_{5(\mu),\lambda\omega}+J^{{\rm dq}(2)}_{5(\mu),\lambda\omega}-J^{{\rm ex}(4)}_{5(\mu),\lambda\omega}\\
-&J^{{\rm sg}(5)}_{5(\mu),\lambda\omega}-
J^{{\rm \overline{sg}}(6)}_{5(\mu),\lambda\omega}\big)\,.
\end{align}
Here, for additional clarity, we have included the diagram label from Fig.\,\ref{figcurrent} as an additional superscript.
In the isospin symmetry limit, expressions for the other $\Delta$ states can be obtained straightforwardly:
\begin{subequations}
\label{VariousJs}
\begin{align}
J^{\Delta^{++}}_{5(\mu),\lambda\omega} &= 3J^{\Delta^+}_{5(\mu),\lambda\omega}\,,\\
J^{\Delta^{0}}_{5(\mu),\lambda\omega} &= -J^{\Delta^+}_{5(\mu),\lambda\omega}\,,\\
J^{\Delta^{-}}_{5(\mu),\lambda\omega} &= -3J^{\Delta^+}_{5(\mu),\lambda\omega}\,.
\end{align}
\end{subequations}
Thus,
\begin{subequations}
\begin{align}
F^{\Delta^{++}} &= 3F^{\Delta^{+}}\,,\\
F^{\Delta^{0}\phantom{+}} &= -F^{\Delta^{+}}\,,\\
F^{\Delta^{-}\phantom{+}} &= -3F^{\Delta^{+}}\,,	
\end{align}
\end{subequations}
where $F\in\{g_1,g_3,h_1,h_3,\tilde{g},\tilde{h},G_{\pi\Delta\Delta},H_{\pi\Delta\Delta}\}$.


\section{Extraction of the form factors}
\label{secffproj}

Beginning with the expressions for the $\Delta$ axial current, Eqs.\,\eqref{axffsdef}, \eqref{axpscurrent}, one can extract the four axialvector form factors by using the following projection matrices:
\begin{subequations}
\label{axproj}
\begin{align}
\label{axprojs1}
{\mathpzc s}_1 &:= i{\rm tr_D}[J_{5\mu,\lambda\omega}\gamma_5]\hat{Q}_\mu \hat{Q}_\lambda \hat{Q}_\omega\,,\\
\label{axprojss}
{\mathpzc s}_2 &:= i{\rm tr_D}[J_{5\mu,\lambda\lambda}\gamma_5]\hat{Q}_\mu\,,\\
\label{axprojs3}
{\mathpzc s}_3 &:= {\rm tr_D}[J_{5\mu,\lambda\omega}\gamma_\mu^T\gamma_5]\hat{Q}_\lambda \hat{Q}_\omega\,,\\
\label{axprojs4}
{\mathpzc s}_4 &:= {\rm tr_D}[J_{5\mu,\lambda\lambda}\gamma_\mu^T\gamma_5]\,,
\end{align}
\end{subequations}
where the trace is over Dirac indices;
$\hat{Q}_\mu=Q_\mu/\sqrt{Q^2}$,
$\gamma_\mu^T=\gamma_\mu - \gamma\cdot \hat{Q} \hat{Q}_\mu$.  In this case, one has
\begin{subequations}
\label{axprojffs}
\begin{align}
g_1 &= \sum_{i=1}^4c_{1i}{\mathpzc s}_i\,,\,\,\,\,\,\,\,g_3 = \sum_{i=1}^4c_{2i}{\mathpzc s}_i\,,\\
h_1 &= \sum_{i=1}^4c_{3i}{\mathpzc s}_i\,,\,\,\,\,\,\,\,h_3 = \sum_{i=1}^4c_{4i}{\mathpzc s}_i\,,
\end{align}
\end{subequations}
with (${\mathpzc t}=Q^2/[4 m_\Delta^2]$)
{\allowdisplaybreaks
\begin{equation}
\label{axprojcoes}
\begin{array}{ll}
c_{11} = 0 = c_{12}\,, & \\[2ex]
\displaystyle c_{13} = \frac{3(1+2{\mathpzc t})}{8(1+{\mathpzc t})^2}\,,
& \displaystyle c_{14} = \frac{-3}{8(1+{\mathpzc t})}\,,\\[2ex]
\displaystyle c_{21} = \frac{1+4{\mathpzc t}}{4{\mathpzc t}^{3/2}(1+{\mathpzc t})}\,,
& \displaystyle c_{22} = \frac{-1}{2{\mathpzc t}^{3/2}}\,,\\[2ex]
\displaystyle c_{23} = \frac{3(1+2{\mathpzc t})}{8{\mathpzc t}(1+{\mathpzc t})^2}\,,
& \displaystyle c_{24} = \frac{-3}{8{\mathpzc t}(1+{\mathpzc t})}\,,\\[2.5ex]
\displaystyle
c_{31} = 0 = c_{32} \,, & \\[2ex]
\displaystyle c_{33} = \frac{3(5+8{\mathpzc t}[1+{\mathpzc t}])}{16{\mathpzc t}(1+{\mathpzc t})^3}\,,
& \displaystyle c_{34} = \frac{-3(1+2{\mathpzc t})}{8{\mathpzc t}(1+{\mathpzc t})^2}\,,\\[2ex]
\displaystyle c_{41} = \frac{5+4{\mathpzc t}(1+2{\mathpzc t})}{4{\mathpzc t}^{5/2}(1+{\mathpzc t})^2}\,,
& \displaystyle c_{42} = \frac{-(1+4{\mathpzc t})}{4{\mathpzc t}^{5/2}(1+{\mathpzc t})}\,,\\[2ex]
\displaystyle c_{43} = \frac{3(5+8{\mathpzc t}[1+{\mathpzc t}])}{16{\mathpzc t}^2(1+{\mathpzc t})^3}\,,
& \displaystyle c_{44} = \frac{-3(1+2{\mathpzc t})}{8{\mathpzc t}^2(1+{\mathpzc t})^2}\,.
\end{array}
\end{equation}
}

Turning to the $\Delta$-baryon pseudoscalar current, \linebreak Eqs.\,\eqref{psffsdef}, \eqref{axpscurrent}, one can extract the other two form factors of interest.  Using the following projection matrices:
\begin{subequations}
\label{psproj}
\begin{align}
\tilde{{\mathpzc s}}_1 &:= i{\rm tr_D}[J_{5,\lambda\omega}\gamma_5]\hat{Q}_\lambda \hat{Q}_\omega\,,\\
\tilde{{\mathpzc s}}_2 &:= i{\rm tr_D}[J_{5,\lambda\lambda}\gamma_5]\,,\end{align}
\end{subequations}
one has
\begin{align}
\label{psprojffs}
\tilde{g} = \sum_{i=1}^2\tilde{c}_{1i}\tilde{{\mathpzc s}}_i\,,\,\,\,\,\,\,\,\tilde{h} = \sum_{i=1}^2\tilde{c}_{2i}\tilde{{\mathpzc s}}_i\,,
\end{align}
with
\begin{equation}
\label{psprojcoes}
\begin{array}{ll}
\displaystyle \tilde{c}_{11} = \frac{-(1+4{\mathpzc t})}{4{\mathpzc t}(1+{\mathpzc t})}\,, &
\displaystyle \tilde{c}_{12} = \frac{1}{2{\mathpzc t}}\,,\\[2ex]
\displaystyle \tilde{c}_{21} = \frac{-(5+4{\mathpzc t}[1+2{\mathpzc t}])}{4{\mathpzc t}^2(1+{\mathpzc t})^2}\,, &
\displaystyle \tilde{c}_{22} = \frac{1+4{\mathpzc t}}{4{\mathpzc t}^2(1+{\mathpzc t})}\,.
\end{array}
\end{equation}


\section{QCD-kindred framework}
\label{secqcdmodel}
Since being introduced in Refs.\,\cite{Ivanov:1998ms, Hecht:2000xa, Alkofer:2004yf}, the QCD-kindred model for ground-state mesons and baryons that we use herein has been refined in a series of analyses that may be traced from Ref.\,\cite{Segovia:2014aza}.  Consistency between the various Schwinger functions involved is guaranteed through their mutual interplay in the description and prediction of hadron observables.

\subsection{Dressed quark propagator}
\label{subappendixqprop}
The dressed-quark propagator is:
{\allowdisplaybreaks
\begin{subequations}
\begin{align}
\label{Spsigma}
S(p) & =  -i \gamma\cdot p\, \sigma_V(p^2) + \sigma_S(p^2) \\
\label{SpAB}
& = 1/[i\gamma\cdot p\, A(p^2) + B(p^2)]\,.
\end{align}
\end{subequations}
Regarding light-quarks, the wave function renormalisation and dressed-quark mass:
\begin{equation}
\label{ZMdef}
Z(p^2)=1/A(p^2)\,,\;M(p^2)=B(p^2)/A(p^2)\,,
\end{equation}
respectively, receive significant momentum-dependent corrections at infrared momenta \cite{Lane:1974he, Politzer:1976tv, Binosi:2016wcx}: $Z(p^2)$ is suppressed and $M(p^2)$ enhanced.  These features are an expression of emergent hadron mass (EHM) \cite{Binosi:2022djx, Papavassiliou:2022wrb, Ding:2022ows, Ferreira:2023fva}.
}

An efficacious parametrisation of $S(p)$, which exhibits the features described above, has been used extensively in hadron studies -- see, \emph{e.g}., \cite{Burkert:2017djo, Chen:2018nsg, Chen:2019fzn, Lu:2019bjs, Cui:2020rmu}.  It is expressed via
{\allowdisplaybreaks
\begin{subequations}
\label{EqSSSV}
\begin{align}
\bar\sigma_S(x) & =  2\,\bar m \,{\cal F}(2 (x+\bar m^2)) \nonumber \\
& \quad + {\cal F}(b_1 x) \,{\cal F}(b_3 x) \,
\left[b_0 + b_2 {\cal F}(\epsilon x)\right]\,,\label{ssm} \\
\label{svm} \bar\sigma_V(x) & =  \frac{1}{x+\bar m^2}\, \left[ 1 - {\cal F}(2
(x+\bar m^2))\right]\,,
\end{align}
\end{subequations}}
\hspace*{-0.5\parindent}with $x=p^2/\lambda^2$, $\bar m$ = $m/\lambda$,
\begin{equation}
\label{defcalF}
{\cal F}(x)= \frac{1-\mbox{\rm e}^{-x}}{x}  \,,
\end{equation}
$\bar\sigma_S(x) = \lambda\,\sigma_S(p^2)$ and $\bar\sigma_V(x) =
\lambda^2\,\sigma_V(p^2)$.
The mass-scale, $\lambda=0.566\,$GeV, and
parameter values
\begin{equation}
\label{tableA}
\begin{array}{ccccc}
   \bar m& b_0 & b_1 & b_2 & b_3 \\\hline
   0.00897 & 0.131 & 2.90 & 0.603 & 0.185
\end{array}\;,
\end{equation}
associated with Eqs.\,\eqref{EqSSSV} were fixed in analyses of light-meson observables \cite{Burden:1995ve, Hecht:2000xa}.  (In Eq.\ (\ref{ssm}), $\epsilon=10^{-4}$ serves only to decouple the large- and intermediate-$p^2$ domains.)

The dimensionless $u=d$ current-mass in Eq.\,(\ref{tableA}) corresponds to
$m_q=5.08\,{\rm MeV}$ 
and the propagator yields the following Euclidean constituent-quark mass, defined by solving $p^2=M^2(p^2)$:
$M_q^E = 0.33\,{\rm GeV}$.
The ratio $M_q^E/m_q = 65$ is one expression of dynamical chiral symmetry breaking (DCSB), a corollary of emergent hadronic mass, in the parametrisation of $S(p)$.  It highlights the infrared enhancement of the dressed-quark mass function.

The dressed-quark mass function generated by Eqs.\ \eqref{EqSSSV} -- \eqref{tableA} is drawn elsewhere \cite[Fig.\,13]{Chen:2021guo}.  The image demonstrates that, although simple and introduced long beforehand, the parametrisation is a sound representation of contemporary numerical results.

The expressions in Eq.\,\eqref{EqSSSV} ensure dressed-quark confinement via the violation of reflection positivity-- see, \emph{e.g}.\ Ref.\,\cite[Sec.\,5]{Ding:2022ows}.  The same is true of the diquark propagators in Eq.\,\eqref{dqprop}.


\subsection{Diquark amplitude and propagator}
\label{secdq}
Regarding $\Delta$-baryons, it is only necessary to involve isovector-axialvector diquarks \cite{Liu:2022ndb}.  Retaining just the dominant structure, their correlation amplitude is
\begin{align}
\label{dqamp}
\Gamma^{1^+}_\mu(k;K) = ig_{1^+}\gamma_\mu C\vec{t}_f\vec{H}_c
{\mathpzc F}(k^2/\omega_{1^+}^2)\,.
\end{align}
Here,
$K$ is the diquark's total momentum;
$k$ is the relative momentum;
${\mathpzc F}$ is the function in Eq.\,\eqref{defcalF};
$\omega_{1^+}$ is a width parameter, which characterises the diquark's propagation within the baryon, $\omega_{1^+}^2=m_{1^+}^2/2$, where $m_{1^+}=0.89\,$GeV is the diquark mass;
$\vec{H}_c = \{i\lambda_c^7, -i\lambda_c^5,i\lambda_c^2\}$, with $\{\lambda_c^k,k=1,\ldots,8\}$ being Gell-Mann matrices in colour space, expresses the diquarks' colour antitriplet character;
$C=\gamma_2\gamma_4$ is the charge-conjugation matrix;
and
$\vec{t}_f=(t_f^1,t_f^2,t_f^3)$ are the flavour matrices:
\begin{subequations}
\label{dqfm}
\begin{align}
t_f^1 &= \frac{1}{2}(\tau^0+\tau^3)\,,\\
t_f^2 &= \frac{1}{\sqrt{2}}\tau^1\,,\\
t_f^3 &= \frac{1}{2}(\tau^0-\tau^3)\,.
\end{align}
\end{subequations}

The coupling constant, $g_{1^+}$, is determined by the canonical normalisation condition:
\begin{subequations}
\label{bsanor}
\begin{align}
& \rule{1em}{0ex} 2K_\mu = \frac{\partial}{\partial Q_\mu}\Pi(K;Q)\bigg|^{K^2=-m_{1^+}^2}_{Q=K}\,,\\
\Pi& (K,Q) = \tfrac{1}{3} T_{\rho\nu}^K \Pi_{\rho\nu}(K;Q) \,, \\
\nonumber
\Pi_{\rho\nu}& (K,Q) = {\rm tr}_{\rm CDF}\int\frac{d^4k}{(2\pi)^4}\bar{\Gamma}_\rho^{1^+}(k;-K) \\
&\times S(k+Q/2) \Gamma_\nu^{1^+}(k;K)S^{\rm T}(-k+Q/2)\,,
\end{align}
\end{subequations}
where
$T_{\rho\nu}^K = [\delta_{\rho\nu}+ K_\rho K_\nu/m_{1^+}^2]$ and
\begin{equation}
\bar{\Gamma}_\mu^{1^+}(k;K) =C^\dagger\Gamma_\mu^{1^+}(-k;K)C\,.
\end{equation}
Using Eqs.\,\eqref{axdqmass}, \eqref{dqamp}, and\eqref{bsanor}, one finds
\begin{align}
g_{1^+} = 12.7\,.
\end{align}

In order to solve the Faddeev equation, Fig.\,\ref{figFaddeev}, one also needs to specify the diquark propagator:
\begin{align}
\label{dqprop}
{\cal D}^{1^+}_{\mu\nu}(K)=\bigg[\delta_{\mu\nu} +\frac{K_\mu K_\nu}{m_{1^+}^2}\bigg]\frac{1}{m_{1^+}^2}{\mathpzc F}(k^2/\omega_{1^+}^2)\,.
\end{align}


\subsection{$\Delta$ Faddeev amplitude}
\label{secqdqamp}
The solution of the $\Delta(1232)$-baryon Faddeev equation, specified generically by Fig.\,\ref{figFaddeev}, takes the  form:
\begin{subequations}
\label{qdqamp}
\begin{align}
\Psi_{\mu\nu}^\Delta(\ell;P)&=\sum^8_{k=1}{\mathpzc a}^\Delta_k(\ell^2;\ell\cdot P)
{\mathpzc D}^k_{\mu\nu}(\ell;P)\frac{\lambda^0_c}{\sqrt{3}}\vec{s}_f\,,\\
{\mathpzc D}^k_{\mu\nu} &= {\mathpzc S}^k\delta_{\mu\nu}\,,\,\,\,\,k=1,2\,,\\
{\mathpzc D}^k_{\mu\nu} &= i\gamma_5{\mathpzc A}^{k-2}_\mu\ell_\nu^\perp\,,\,\,\,\,k=3,\dots,8\,,
\end{align}
\end{subequations}
where
\begin{subequations}
\begin{align}
{\mathpzc S}^1 &= {\mathbb I}_{\rm D}\,,\\
{\mathpzc S}^2 &= i\gamma\cdot\hat{\ell}-\hat{\ell}\cdot\hat{P} {\mathbb I}_{\rm D}\,,\\
{\mathpzc A}^{1}_\mu &= \gamma\cdot\ell^\perp\hat{P}_\mu\,,\\
{\mathpzc A}^{2}_\mu &= -i\hat{P}_\mu{\mathbb I}_{\rm D}\,,\\
{\mathpzc A}^{3}_\mu &= \gamma\cdot\hat{\ell}^\perp \hat{\ell}^\perp_\mu\,,\\
{\mathpzc A}^{4}_\mu &= i\hat{\ell}^\perp_\mu {\mathbb I}_{\rm D}\,,\\
{\mathpzc A}^{5}_\mu &= \gamma_\mu^\perp - {\mathpzc A}^{3}_\mu\,,\\
{\mathpzc A}^{6}_\mu &= i\gamma_\mu^\perp\gamma\cdot\hat{\ell}^\perp - {\mathpzc A}^{4}_\mu\,,
\end{align}
\end{subequations}
are the Dirac basis matrices, with $\hat{\ell}^2=1$, $\hat{P}^2=-1$, $\ell^\perp_\nu=\hat{\ell}_\nu+\hat{\ell}\cdot \hat{P}\hat{P}_\nu$, $\gamma^\perp_\nu=\gamma_\nu+\gamma\cdot\hat{P}\hat{P}_\nu$;
$\lambda_c^0 ={\rm diag}[1,1,1]$ is a colour matrix;
and $\vec{s}_f$ are the flavour matrices of the quark+diquark amplitude, which are obtained by removing the diquark's flavour matrices \eqref{dqfm} from the $\Delta$'s full amplitude,
\begin{subequations}
\label{qdqfm}
\begin{align}
s_f^1 &= \left(\begin{array}{cccc} 1 & 0 & 0 & 0\\ 0 & \sqrt{\frac{1}{3}} & 0 & 0\end{array}\right)\,,\\
s_f^2 &= \left(\begin{array}{cccc} 0 & \sqrt{\frac{2}{3}} & 0 & 0\\ 0 & 0 & \sqrt{\frac{2}{3}} & 0\end{array}\right)\,,\\
s_f^3 &= \left(\begin{array}{cccc} 0 & 0 & \sqrt{\frac{1}{3}} & 0\\ 0 & 0 & 0 & 1\end{array}\right)\,.
\end{align}
\end{subequations}

Upon solving the Faddeev equation, one obtains all scalar functions in Eq.\,\eqref{qdqamp} and the $\Delta$-baryon mass.  Using Eq.\,\eqref{axdqmass},
\begin{equation}
\label{massDelta}
m_\Delta=1.35\,{\rm GeV}\,.
\end{equation}
Notably, the kernel in Fig.\,\ref{figFaddeev} omits all those contributions which may be linked with meson-baryon final-state interactions, \emph{i.e}., the terms resummed in DCC models in order to transform a bare-baryon into the observed state \cite{JuliaDiaz:2007kz, Suzuki:2009nj, Ronchen:2012eg, Kamano:2013iva}.
The Faddeev equation outputs should thus be viewed as describing the \emph{dressed-quark core} of the $\Delta$-baryon, not the completely-dressed, observable object \cite{Eichmann:2008ae, Eichmann:2008ef, Roberts:2011cf}.
In support of this interpretation, we refer to Ref.\,\cite[Fig.\,4]{Liu:2022ndb}, which shows mass predictions for the four lowest-lying $\Delta$-baryon multiplets.  Evidently, by subtracting $\delta_{\rm MB}=0.17\,$GeV from each calculated mass, a value that matches the offset between bare and dressed $\Delta(1232)\tfrac{3}{2}^+$ masses determined in the DCC analysis of Ref.\,\cite{Suzuki:2009nj}, one finds level orderings and splitting that  match well with experiment.


\section{Current diagrams}
\label{seccurr}
In Fig.\,\ref{figcurrent}, we draw the symmetry preserving current appropriate to a bar\-yon whose structure is prescribed by the Faddeev equation indicated by Fig.\,\ref{figFaddeev}.  In general there are six distinct sorts of terms.  Diagrams (1) and (2) may be called impulse contributions: the probe strikes either a quark or a diquark.
Diagram (3) is a partner to Diagram (2).  In cases where more than one type of diquark correlation is present in the target baryon, then this contribution expresses probe-induced transitions between different types, \emph{e.g}., in the nucleon, it describes transitions between scalar and axialvector diquarks.
Naturally, since there are only axialvector diquarks in $\Delta$-baryons, this diagram vanishes in the calculation of $\Delta$-baryon elastic form factors.  That is why it is ``red boxed'' in Fig.\,\ref{figcurrent}.
Impulse contributions are typically one-loop diagrams, \emph{i.e}., four dimensional integrals; and when that is the case, they can readily be evaluated using Gaussian quadrature methods.

The remaining contributions appear because the \linebreak quark exchanged in the Faddeev equation kernel is also struck by the probe.  Diagram (4) is the explicit interaction contribution.  Diagrams (5), (6) are so-called seagull terms, whose presence guarantees that all Ward-Green-Takahashi identities associated with the interaction probe are preserved at the baryon level.  The seagulls for electromagnetic interactions were derived in Ref.\,\cite{Oettel:1999gc} and those for weak interactions in Ref.\,\cite{Chen:2021guo}.
These three contributions are two-loop diagrams, which we evaluate using Monte-Carlo methods.

For explicit calculations, we use the Breit frame: $P_i=K-Q/2$, $P_f=K+Q/2$, $K=(0,0,0,i E_\Delta(Q))$, $E_\Delta^2(Q) = m_\Delta^2+Q^2/4$.


\subsection{Diagram (1)}
\label{secdia1}
Probe coupling directly to the uncorrelated quark:
\begin{align}
\label{dia1}
\nonumber
&J^{\rm q}_{5(\mu),\lambda\omega}(K,Q) = \int_{dp} \bar{\Psi}^\Delta_{\lambda\alpha}(p'_f;-P_f)S(p_{q_+})\\
&\quad \times \Gamma^j_{5(\mu)}(p_{q+},p_{q-})S(p_{q-}){\cal D}^{1^+}_{\alpha\beta}(p_d)\Psi^\Delta_{\beta\omega}(p'_i;P_i)\,,
\end{align}
where
$\Gamma^j_{5(\mu)}$ is the dressed-quark pseudoscalar (axialvector) vertex,
\begin{subequations}
\label{D1momenta}
\begin{align}
p^\prime_i & = p - \hat{\eta}Q/2\,, & p'_f & = p + \hat{\eta}Q/2\,,\\
p_{q-} & = p^\prime_i + \eta P_i\,, &  p_{q+} & = p^\prime_f + \eta P_f\,,
\end{align}
\end{subequations}
$p_d  = \hat{\eta}P_i - p^\prime_i = \hat{\eta}P_f - p^\prime_f$.

The dressed-quark axialvector and pseudoscalar vertices satisfy the axialvector Ward-Green-Takahashi identity (AWGTI):
\begin{align}
\label{qaxwti}
\nonumber
Q_\mu\Gamma^j_{5\mu}&(k_+,k_-)  +2im_q	\Gamma^j_{5}(k_+,k_-)\\
=& S^{-1}(k_+)i\gamma_5\frac{\tau^j}{2}+\frac{\tau^j}{2}i\gamma_5S^{-1}(k_-)\,,
\end{align}
where $Q$ is the incoming probe momentum, $k_-$, $k_+$ are the incoming and outgoing quark momenta, $k_\pm = k\pm Q/2$.
Preserving the AWGTI is crucial for PCAC \cite{Chen:2021guo} and the following forms ensure this outcome:
\begin{subequations}
\label{axpsvtx}
\begin{align}
\label{axvtx}
\nonumber
\Gamma^j_{5\mu}(k_+,k_-) & = \frac{\tau^j}{2}\gamma_5\bigg[\gamma_\mu\Sigma_A^{+-}
+ 2\gamma\cdot k k_\mu\Delta_A^{+-}\\
&\quad +2i\frac{Q_\mu}{Q^2+m_\pi^2}\Sigma_B^{+-}\bigg]\,,\\
\label{psvtx}
i\Gamma^j_{5}(k_+,k_-) & = \frac{\tau^j}{2} \frac{m_\pi^2}{Q^2+m_\pi^2}
\frac{1}{m_q}i\gamma_5\Sigma_B^{+-}\,,
\end{align}
\end{subequations}
where
\begin{subequations}
\begin{align}
\Sigma_F^{+-} &= \frac{1}{2}[F(k_+^2)+F(k_-^2)]\,,\\
\Delta_F^{+-} &= \frac{F(k_+^2)-F(k_-^2)}{k_+^2-k_-^2}\,,
\end{align}
\end{subequations}
with $F\in\{A,B\}$ and $A$ and $B$ are the dressing functions in the quark propagator -- see Eq.\,\eqref{SpAB} in \ref{subappendixqprop}.


\subsection{Diagram (2)}
\label{secdia2}

Probe coupling to an axialvector diquark:
\begin{align}
\label{dia2}
\nonumber
J^{\rm dq}_{5(\mu),\lambda\omega}(K,Q) = &\int_{dp}\bar{\Psi}^\Delta_{\lambda\alpha}(p''_f;-P_f){\cal D}^{1^+}_{\alpha\rho}(p_{d+})\\
\nonumber
&\times \Gamma_{5(\mu),\rho\sigma}^{AA}(p_{d+},p_{d-}){\cal D}^{1^+}_{\sigma\beta}(p_{d-})\\
&\times S(p_q)\Psi^\Delta_{\beta\omega}(p''_i;P_i)\,,
\end{align}
where $\Gamma_{5(\mu),\rho\sigma}^{AA}$ is the axialvector diquark pseudoscalar (axialvector) vertex and
\begin{subequations}
\begin{align}
p^{\prime\prime}_i &= p + \eta Q/2\,, & p^{\prime\prime}_f &= p - \eta Q/2\,,\\
p_{d-} &= \hat{\eta}P_i - p^{\prime\prime}_i\,, & p_{d+} &= \hat{\eta}P_f - p^{\prime\prime}_f\,,
\end{align}
\end{subequations}
$p_q = p^{\prime\prime}_i +\eta P_i = p^{\prime\prime}_f +\eta P_f$.

\begin{figure}[t]
\centerline{%
\includegraphics[clip, width=0.47\textwidth]{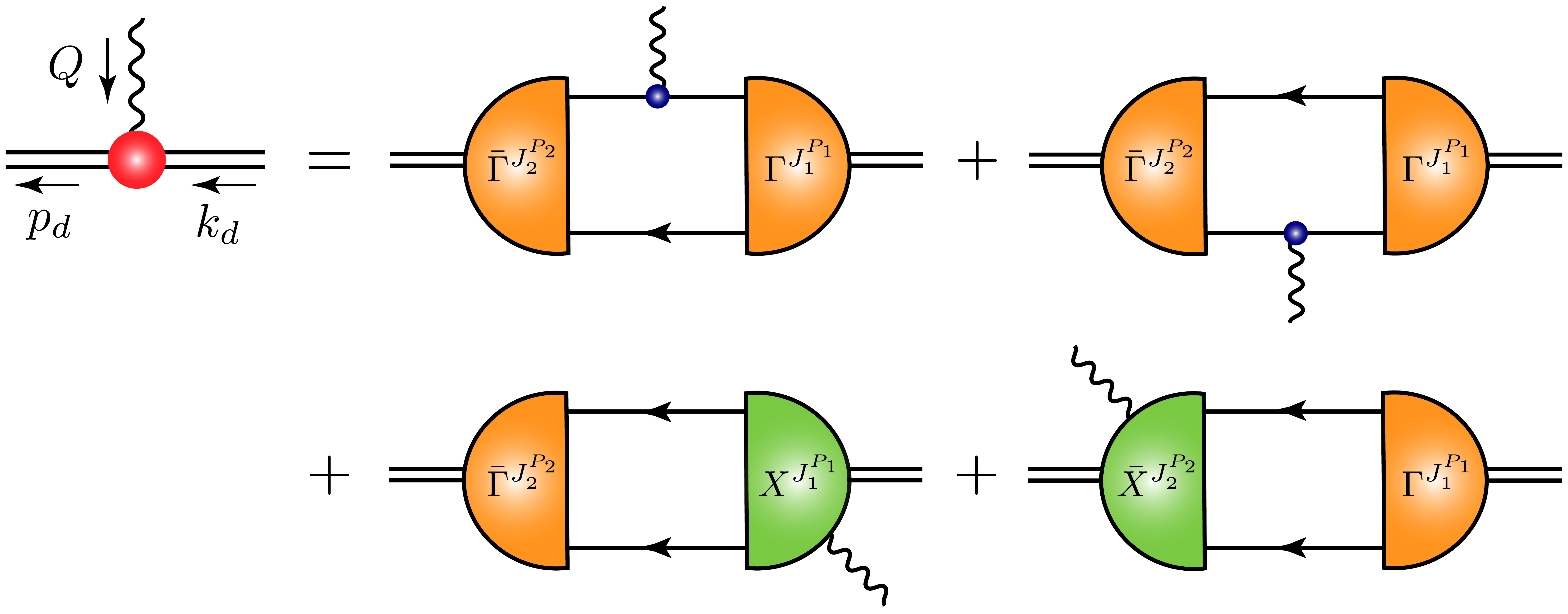}}
\caption{\label{figdqvx}
Interaction vertex for the $J_1^{P_1}\to J_2^{P_2}$ diquark transition ($p_d=k_d+Q$): \emph{single line}, quark propagator; \emph{undulating  line}, the axial or pseudoscalar current; $\Gamma$, diquark correlation amplitude; \emph{double line}, diquark propagator; and $\chi$, seagull interaction.
For $\Gamma_{5(\mu),\rho\sigma}^{AA}$ in Eq.\,\eqref{dia2}, $J_1^{P_1}= J_2^{P_2} = 1^+$.
}
\end{figure}

Each current-diquark vertex receives four contributions, \emph{viz}.\ those depicted in Fig.\,\ref{figdqvx}.
Two of them are generated by coupling the current to the upper and lower quark lines of the resolved diquark.
The remaining two are current couplings to the diquark amplitudes, \emph{i.e.}, ``seagull terms'' -- see \ref{secdia56} for details.
Consequently, $\Gamma_{5(\mu),\rho\sigma}^{AA}$ can be expressed as the following one-loop integral:
\begin{align}
\label{dqvxrev}
\nonumber
& \Gamma_{5(\mu),\rho\sigma}^{AA}(p_d,k_d) = \\
\nonumber
-&{\rm tr_D}\int_{d\ell}\bigg[{\bar\Gamma}^{1^+}_\rho(\ell_r)S(\ell_2)\Gamma^{j}_{5(\mu)}(\ell_2,\ell_1)S(\ell_1)\Gamma^{1^+}_\sigma(\ell'_r)S^{\rm T}(\ell_3)\\
\nonumber
+&\bar{\Gamma}^{1^+}_\rho(\tilde{\ell}_r)S(\tilde{\ell}_3)\Gamma^{1^+}_\sigma(\tilde{\ell}'_r)\big[S(\tilde{\ell}_2)\Gamma^j_{5(\mu)}(\tilde{\ell}_2,\tilde{\ell}_1)S(\tilde{\ell}_1)\big]^{\rm T}\\
\nonumber
+& \bar{\Gamma}^{1^+}_\rho(\ell)S(\ell_+)\chi^{1^+}_{5(\mu),\sigma}(\ell,Q)S^{\rm T}(-\ell_-)\\
+& \bar{\chi}^{1^+}_{5(\mu),\rho}(\ell,Q)
S(\tilde{\ell}_+)\Gamma^{1^+}_\sigma(\ell)S^{\rm T}(-\tilde{\ell}_-)\bigg]\,,
\end{align}
with
{\allowdisplaybreaks
\begin{subequations}
\begin{align}
\ell_{^1_2} &= \frac{p_d+k_d}{4} \mp \frac{Q}{2}+\ell\,,\\
\ell_3 &= \frac{p_d+k_d}{4} - \ell\,,\\
\ell_r &= \frac{\ell_2-\ell_3}{2}\,,
\ell^\prime_r = \frac{\ell_1-\ell_3}{2}\,,\\
\tilde{\ell}_{^1_2} &= \frac{p_d+k_d}{4}\mp\frac{Q}{2}-\ell\,,\\
\tilde{\ell}_3 &= \frac{p_d+k_d}{4} + \ell\,,\\
\tilde{\ell}_r &= \frac{\tilde{\ell}_2-\tilde{\ell}_3}{2}\,,
\tilde{\ell}^\prime_r = \frac{\tilde{\ell}_1-\tilde{\ell}_3}{2}\,,\\
\ell_\pm & = \frac{p_d}{2} \pm \ell\,,
\tilde{\ell}_\pm = \frac{k_d}{2} \pm \ell\,.
\end{align}
\end{subequations}
}

Inserting Eq.\,\eqref{dqvxrev} into Eq.\,\eqref{dia2}, it becomes clear that Diagram (2) is, herein, a two-loop diagram, and its computation requires Monte-Carlo methods.

Notably, for nucleon axial form factors, Refs.\,\cite{Chen:2020wuq, Chen:2021guo, ChenChen:2022qpy} constructed \emph{Ans\"atze} for the current-diquark vertices, ensuring that Diagram (2) remained a 1-loop integral.
This approach cannot efficiently be employed herein because the $\Delta$-baryon has two independent sets of axialvector and pseudoscalar form factors, \emph{viz}.\ $\{g_1,g_3,\tilde{g}\}$ and $\{h_1,h_3,\tilde{h}\}$.


\subsection{Diagram (4)}
\label{secdia4}

Probe coupling to the quark exchanged as one diquark breaks-up and another is formed:
\begin{align}
\label{dia4}
\nonumber
&J^{\rm ex}_{5(\mu),\lambda\omega}(K,Q) = \int_{dp}\int_{dk} \bar{\Psi}^\Delta_{\lambda\rho}(p;-P_f){\cal D}^{1^+}_{\rho\alpha}(\tilde{p}_{d+})\\
\nonumber
\times &S(\tilde{p}_{q+})\Gamma^{1^+}_\alpha(\tilde{k}_r)\big[S(\tilde{q}')\Gamma^j_{5(\mu)}(\tilde{q}',\tilde{q})S(\tilde{q})\big]^{\rm T}\bar{\Gamma}^{1^+}_\beta(\tilde{p}'_r)\\
\times & S(\tilde{p}_{q-}){\cal D}^{1^+}_{\beta\sigma}(\tilde{p}_{d-})
\Psi^\Delta_{\sigma\omega}(k;P_i)\,,
\end{align}
with
\begin{subequations}
\begin{align}
\tilde{p}_{q-} &= k +\eta P_i\,, \;
\tilde{p}_{q+} = p +\eta P_f\,,\\
\tilde{p}_{d-} &= \hat{\eta}P_i - k\,, \;
\tilde{p}_{d+} = \hat{\eta}P_f - p\,,\\
\tilde{k}_r &= \frac{1}{2}((k+\hat{\eta}\,Q)+2p+(3\eta-1)P_f)\,,\\
\tilde{p}_r &= \frac{1}{2}(p+2(k+\hat{\eta}\,Q)+(3\eta-1)P_f)\,,\\
\tilde{q} &= -p-(k+\hat{\eta}\,Q)+(1-2\eta)P_f\,,\\
\tilde{k}^\prime_r &= \tilde{k}_r -Q\,,\;
\tilde{p}^\prime_r = \tilde{p}_r -Q\,,\;
\tilde{q}^\prime = \tilde{q} +Q\,.
\end{align}
\end{subequations}
The process of quark exchange in their Faddeev kernel provides the attraction required to bind the $\Delta$-baryon.  It also ensures that the Faddeev amplitude has the correct antisymmetry under the exchange of any two dressed quarks. These features are absent in models with pointlike diquarks.


\subsection{Diagrams (5) and (6)}
\label{secdia56}

Owing to the nonpointlike character of the diquark correlations, one must also consider couplings of the incoming probe to the diquark amplitudes, \emph{viz}.\ ``seagull terms'', which appear as partners to Diagram (4) and are necessary to ensure current conservation \cite{Oettel:1999gc}.  The seagull terms for the axialvector and pseudoscalar currents are derived in Ref.\,\cite{Chen:2021guo}. One has
{\allowdisplaybreaks
\begin{align}
\nonumber
&J^{\rm sg}_{5(\mu),\lambda\omega}(K,Q) = \int_{dp}\int_{dk} \bar{\Psi}^\Delta_{\lambda\rho}(p;-P_f){\cal D}^{1^+}_{\rho\alpha}(\tilde{p}_{d+})\\
\nonumber
& \quad \times S(\tilde{p}_{q+})\chi^{j,1^+}_{5(\mu),\alpha}(k_1,Q)S^{\rm T}(\tilde{q}')\bar{\Gamma}^{1^+}_{\beta}(\tilde{p}'_r)S(\tilde{p}_{q-})\\
& \quad \times  {\cal D}^{1^+}_{\beta\sigma}(\tilde{p}_{d-})
\Psi^\Delta_{\sigma\omega}(k;P_i)\,,
\end{align}
for Diagram (5), and
\begin{align}
\nonumber
&J^{\overline{\rm sg}}_{5(\mu),\lambda\omega}(K,Q) = \int_{dp}\int_{dk} \bar{\Psi}^\Delta_{\lambda\rho}(p;-P_f){\cal D}^{1^+}_{\rho\alpha}(\tilde{p}_{d+})\\
\nonumber
& \quad \times S(\tilde{p}_{q+})\Gamma^{1^+}_\alpha(\tilde{k}_r)S^{\rm T}(\tilde{q})\bar{\chi}^{j,1^+}_{5(\mu),\beta}(k_2,Q)S(\tilde{p}_{q-})\\
& \quad \times  {\cal D}^{1^+}_{\beta\sigma}(\tilde{p}_{d-})
\Psi^\Delta_{\sigma\omega}(k;P_i)\,,
\end{align}
for Diagram (6).  The momenta are
\begin{align}
k_1 &= \frac{\tilde{p}_{q+} -\tilde{q}'}{2}\,, \;
k_2 = \frac{\tilde{p}_{q-} -\tilde{q}}{2}\,.
\end{align}
Explicitly, the seagull terms are:
\begin{subequations}
\label{sg}
\begin{align}
\nonumber
\label{axsg}
\chi^{j,1^+}_{5\mu,\alpha}(k,Q) = &-\frac{Q_\mu}{Q^2+m_\pi^2}\bigg[\frac{\tau^j}{2}i\gamma_5\Gamma^{1^+}_\alpha(k-Q/2)+\\
&\Gamma^{1^+}_\alpha(k+Q/2)(i\gamma_5\frac{\tau^j}{2})^{\rm T}\bigg]\,, \\
\nonumber
\label{pssg}
i\chi^{j,1^+}_{5,\alpha}(k,Q) = &-\frac{1}{2m_q}\frac{m_\pi^2}{Q^2+m_\pi^2}\bigg[\frac{\tau^j}{2}i\gamma_5\Gamma^{1^+}_\alpha(k-Q/2)+\\
&\Gamma^{1^+}_\alpha(k+Q/2)(i\gamma_5\frac{\tau^j}{2})^{\rm T}\bigg]\,,\\
\nonumber
\label{axsgb}
\bar{\chi}^{j,1^+}_{5\mu,\alpha}(k,Q) = &-\frac{Q_\mu}{Q^2+m_\pi^2}\bigg[\bar{\Gamma}^{1^+}_\alpha(k+Q/2)\frac{\tau^j}{2}i\gamma_5+\\
&(i\gamma_5\frac{\tau^j}{2})^{\rm T}\bar{\Gamma}^{1^+}_\alpha(k-Q/2)\bigg]\,, \\
\nonumber
\label{pssgb}
i\bar{\chi}^{j,1^+}_{5,\alpha}(k,Q) = &-\frac{1}{2m_q}\frac{m_\pi^2}{Q^2+m_\pi^2}\bigg[\bar{\Gamma}^{1^+}_\alpha(k+Q/2)\frac{\tau^j}{2}i\gamma_5+\\
&(i\gamma_5\frac{\tau^j}{2})^{\rm T}\bar{\Gamma}^{1^+}_\alpha(k-Q/2)\bigg]\,.
\end{align}
\end{subequations}
}

The axialvector and pseudoscalar seagull terms are related by the following AWGTI:
\begin{align}
\label{sgaxwti}
\nonumber
&Q_\mu \chi^{j,1^+}_{5\mu,\alpha}(k,Q) + 2im_q \chi^{j,1^+}_{5,\alpha}(k,Q)\\
=& -\frac{\tau^j}{2}i\gamma_5 \Gamma^{1^+}_\alpha(k-Q/2)-
\Gamma^{1^+}_\alpha(k+Q/2)(i\gamma_5\frac{\tau^j}{2})^{\rm T}\,.
\end{align}
Using Eqs.\,\eqref{qaxwti}, \eqref{sgaxwti}, it is straightforward to obtain the AWGTI for the axialvector diquark vertices, Eq.\,\eqref{dqvxrev}:
\begin{align}
Q_\mu \Gamma^{AA}_{5\mu,\rho\sigma}(p_d,k_d)+2im_q \Gamma^{AA}_{5,\rho\sigma}(p_d,k_d)=0\,.
\end{align}


\section{Interpolation of the form factors}
\label{secintpl}

\begin{table}[!t]
\caption{\label{tablepade}
Interpolation parameters for $\Delta$-baryon's axial-vector and $\pi$-$\Delta$ form factors, Eqs.\eqref{axpade} and \eqref{pspade}.
}
\begin{center}
\begin{tabular*}
{\hsize}
{
l@{\extracolsep{0ptplus1fil}}
|l@{\extracolsep{0ptplus1fil}}
l@{\extracolsep{0ptplus1fil}}
l@{\extracolsep{0ptplus1fil}}
l@{\extracolsep{0ptplus1fil}}}\hline
 & $a_0$ & $\phantom{-}a_1$ & $b_1$ & $b_2$ \\\hline
 $g_1$ & $\phantom{1}0.71$ & $\phantom{-}0.024$ & $1.25$ & $0.56$ \\
 $g_3$ & $\phantom{1}1.00$ & $\phantom{-}0.046$ & $1.23$ & $0.60$ \\
 $h_1$ & $\phantom{1}2.40$ & $-0.32$ & $1.30$ & $0.52$ \\
 $h_3$ & $\phantom{1}3.43$ & $-0.47$ & $1.31$ & $0.52$ \\  \hline
 $G_{\pi\Delta\Delta}$ & $10.16$ & $-2.55$ & $1.29$ & $1.14$ \\
 $H_{\pi\Delta\Delta}$ & $34.50$ & $-7.64$ & $1.62$ & $1.22$ \\  \hline
\end{tabular*}
\end{center}
\end{table}

For $-m_\pi^2<s=Q^2<1.6\,m_\Delta^2$, the $\Delta$-baryon axialvector and $\pi$-$\Delta$ form factors can accurately be interpolated using the following function:
\begin{align}
\label{axpade}
\frac{a_0 + a_1s }{1 + b_1s + b_2s^2}\,,
\end{align}
for $g_1$, $h_1$, $G_{\pi\Delta\Delta}$ and $H_{\pi\Delta\Delta}$; and
\begin{align}
\label{pspade}
\frac{a_0 + a_1s }{1 + b_1s + b_2s^2}{\mathpzc R}(s)\,,
\end{align}
for $g_3$ and $h_3$, where
\begin{align}
\label{EqR}
{\mathpzc R}(s)=\frac{m_\pi^2}{s+m_\pi^2}\,\frac{m_\Delta}{f_\pi}\,.
\end{align}
The coefficients for the central results are listed in Table\,\ref{tablepade}.


\end{document}